\documentclass[12pt]{article}
\pdfoutput=1
\usepackage[nosort]{cite}

\usepackage{draft,hyperref,tikz,float,subfig,cite,ifthen}

\usepackage{amsmath}

\makeatletter
\newcommand\xleftrightarrow[2][]{%
  \ext@arrow 9999{\longleftrightarrowfill@}{#1}{#2}}
\newcommand\longleftrightarrowfill@{%
  \arrowfill@\leftarrow\relbar\rightarrow}
\makeatother

\newcommand{\act}[4]{
\begin{gathered}
\begin{tikzpicture}[scale = #1]
\tikzset{line/.style={line width=0.25mm},
curve/.style={line,smooth,tension=1}}
\filldraw (0,0) circle (.05);
\node at (0,-.5) {$\varphi(x)$};
\ifthenelse{\equal{#3}{2}}{
	\draw [line] (0,0) -- (0,2);
}{
	\ifthenelse{\equal{#3}{1}}{
		\draw [line,dashed] (0,0) -- (0,.7);
		\draw [line,dashed] (0,1.3) -- (0,2);
		\draw [line] (0,.7) -- (0,1.3);
	}{}
};
\ifthenelse{\equal{#4}{1}}{
	\ifthenelse{\equal{#2}{2}}{
		\draw [line] (0,0) ++(-240:1) arc (-240:60:1);
		\draw [curve] plot coordinates {(-.5,.866) (-.25,1.05) (0,1.3)};
		\draw [curve] plot coordinates {(.5,.866) (.25,.875) (0,.7)};
	}{
		\ifthenelse{\equal{#2}{1}}{
			\draw [line,dashed] (0,0) ++(-240:1) arc (-240:60:1);
			\draw [curve,dashed] plot coordinates {(-.5,.866) (-.25,1.05) (0,1.3)};
			\draw [curve,dashed] plot coordinates {(.5,.866) (.25,.9) (0,.7)};
		}{}
	}
}{
	\ifthenelse{\equal{#2}{2}}{
		\draw [line] (0,0) ++(-240:1) arc (-240:60:1);
		\draw [curve] plot coordinates {(.5,.866) (.25,1.05) (0,1.3)};
		\draw [curve] plot coordinates {(-.5,.866) (-.25,.875) (0,.7)};
	}{
		\ifthenelse{\equal{#2}{1}}{
			\draw [line,dashed] (0,0) ++(-240:1) arc (-240:60:1);
			\draw [curve,dashed] plot coordinates {(.5,.866) (.25,1.05) (0,1.3)};
			\draw [curve,dashed] plot coordinates {(-.5,.866) (-.25,.9) (0,.7)};
		}{}
	}
}
\end{tikzpicture}
\end{gathered}
}

\newcommand{\Zthree}{
\tikzset{line/.style={line width=0.25mm},
curve/.style={line,smooth,tension=1}}
\draw [line] (0,0) -- (3,0) -- (3,3) -- (0,3) -- (0,0);
\draw [line] (0,1) -- (3,1);
\draw [line] (0,2) -- (3,2);
}

\newcommand{\Zpre}[5]{
\tikzset{line/.style={line width=0.25mm},
curve/.style={line,smooth,tension=1}}
\draw [line] (0,0) -- (2,0) -- (2,2) -- (0,2) -- (0,0);
\ifthenelse{\equal{#4}{}}{
	\ifthenelse{\equal{#2}{1}}{\draw [line,dashed] (0,1) -- (2,1)}{
		\ifthenelse{\equal{#2}{2}}{\draw [line] (0,1) -- (2,1)}{}
	};
	\ifthenelse{\equal{#3}{1}}{\draw [line,dashed] (1,0) -- (1,2)}{
		\ifthenelse{\equal{#3}{2}}{\draw [line] (1,0) -- (1,2)}{}
	};
}{
	\ifthenelse{\equal{#3}{1}}{
			\draw [line,dashed] (1,0) -- (1,0.7);
			\draw [line,dashed] (1,2) -- (1,1.3);
			\draw [line] (1,0.7) -- (1,1.3);
			
		}{
		\ifthenelse{\equal{#3}{2} \AND \not\equal{#2}{2}}{\draw [line] (1,0) -- (1,2)}{
			\ifthenelse{\equal{#3}{2} \AND \equal{#2}{2}}{
				\draw [line] (1,0) -- (1,0.7);
				\draw [line] (1,2) -- (1,1.3);
				\ifthenelse{\equal{#5}{1}}{
					\draw [line,dashed] (1,0.7) -- (1,1.3);
				}{};
			}{};
		};
	};
	\ifthenelse{\equal{#4}{1}}{
		\ifthenelse{\equal{#2}{1}}{
			\draw [curve,dashed] plot coordinates {(0,1) (0.7,1) (1,1.3)};
			\draw [curve,dashed] plot coordinates {(2,1) (1.3,1) (1,0.7)};
			}{
			\ifthenelse{\equal{#2}{2}}{
			\draw [curve] plot coordinates {(0,1) (0.7,1) (1,1.3)};
			\draw [curve] plot coordinates {(2,1) (1.3,1) (1,0.7)};
			}{};
		};
	}{
		\ifthenelse{\equal{#4}{0}}{
			\ifthenelse{\equal{#2}{1}}{
				\draw [curve,dashed] plot coordinates {(0,1) (0.7,1) (1,0.7)};
				\draw [curve,dashed] plot coordinates {(2,1) (1.3,1) (1,1.3)};
				}{
				\ifthenelse{\equal{#2}{2}}{
				\draw [curve] plot coordinates {(0,1) (0.7,1) (1,0.7)};
				\draw [curve] plot coordinates {(2,1) (1.3,1) (1,1.3)};
				}{};
			};
		};
	};
};
}

\newcommand{\Z}[5]{
\begin{gathered}
\begin{tikzpicture}[scale = #1]
\Zpre{#1}{#2}{#3}{#4}{#5}
\end{tikzpicture}
\end{gathered}
}

\begin{document}

\begin{titlepage}

\preprint{CALT-TH-2019-043}

\begin{center}

\hfill \\
\hfill \\
\vskip 1cm

\title{
Duality Defect of the Monster CFT
}

\author{Ying-Hsuan Lin$^{a}$ and Shu-Heng Shao$^b$}

\address{${}^a$Walter Burke Institute for Theoretical Physics,\\ California Institute of Technology,
Pasadena, CA 91125, USA}
\address{${}^b$School of Natural Sciences, Institute for Advanced Study,\\
Princeton, NJ 08540, USA}

\email{yhlin@caltech.edu, shao@ias.edu}

\end{center}

\vfill

\abstract{
We show that the fermionization of the Monster CFT with respect to $\mathbb{Z}_{2A}$ is the tensor product of a free fermion and the Baby Monster CFT.  
The chiral fermion parity of the free fermion implies that the Monster CFT is self-dual under the $\mathbb{Z}_{2A}$ orbifold, {\it i.e.} it enjoys the Kramers-Wannier duality.  
The Kramers-Wannier duality defect extends the Monster group to a larger category of topological defect lines that contains an Ising subcategory. 
We introduce the \textit{defect McKay-Thompson series} defined as the Monster partition function twisted by the  duality defect, and find that   the coefficients can be decomposed into the dimensions of the (projective)   irreducible representations  of the Baby Monster group.  
 We further prove that the defect McKay-Thompson series is invariant under the genus-zero congruence subgroup $16D^0$ of $PSL(2,\mathbb{Z})$.
}

\vfill

\end{titlepage}

\eject

\tableofcontents

\section{Introduction}

The Monster CFT \cite{Frenkel3256}  is a  $c=24$ holomorphic CFT without Kac-Moody symmetry.  
As the name suggests, it has an extremely rich global symmetry group, the Monster group, which is the largest sporadic finite simple group. 
The torus partition function of the Monster CFT is $j(\tau) - 744$, whose coefficients exhibit the Monstrous Moonshine \cite{Thompson,Conway}, as originally observed by John McKay. Hence the very existence of the Monster CFT provides a physical explanation for aspects of the moonshine phenomenon.  
Monster CFT has also been proposed as a  possible holographic dual for ``pure'' gravity in AdS$_3$ \cite{Witten:2007kt}. 

The Monster group has two order two conjugacy classes, $2A$ and $2B$.  For each class, we pick one representative   and denote it as $\bZ_{2A}$ and $\bZ_{2B}$, respectively. 
Both $\bZ_{2A}$ and $\bZ_{2B}$ are non-anomalous, and their orbifolds give the Monster CFT and the Leech lattice CFT, respectively.

\begin{figure}[t]
\centering
\begin{tikzpicture}
\coordinate (Z2)  at (0,0);
\draw (4,0) node {Monster};
\draw (-1,0) node {Baby $\otimes$ MW};
\draw (4,-3) node {Leech};
\draw (9,0) node {B$\&$B};
\draw (9.5,-3) node {B$\&$B $\otimes$ $(-1)^{\text{Arf}}$};
\draw [->,thick] (0.6,0.1)--(2.5,0.1) node[midway,above] {bosonize};
\draw [<-,thick] (0.6,- 0.1)--(2.5,-0.1) node[midway,below] {fermionize $2A$};
\draw [<-,thick] (5.5,0.1)--(7.4,0.1) node[midway,above] {bosonize};
\draw [->,thick] (5.5,- 0.1)--(7.4,-0.1) node[midway,below] {fermionize $2B$};
\draw  [<->, thick]  (4,-.6) -- node[left] {gauge $\mathbb{Z}_{2B}$} (4,-2.5) ;
\draw  [<->, thick]  (9,-.6) -- node[right] {$\otimes~(-1)^{\text{Arf}}$} (9,-2.5);
\draw [<-,thick] (5.5,0.1-3) -- (7.4,0.1-3) node[midway,above] {bosonize};
\draw [->,thick] (5.5,- 0.1 - 3)--(7.4,-0.1-3) node[midway,below] {fermionize $2B$};
\draw  (-1,2) node {$\otimes(-1)^{\text{Arf}}$};
\draw  (4,2) node {gauge $\mathbb{Z}_{2A}$};
\draw[ ->] (-1,1) ++ (240:.5) arc (240:-60:.5);
\draw[ ->] (4,1) ++ (240:.5) arc (240:-60:.5);
\end{tikzpicture}
\caption{The bosonization/fermionization of the Monster CFT with respect to the $\bZ_{2A}$ and the $\bZ_{2B}$ symmetries. Here ``B$\&$B" stands for the ``Beauty and the Beast" ${\cal N}=1$ SCFT of \cite{Dixon:1988qd}, while ``Baby" and ``MW" stand for the Baby Monster CFT and the Majorana-Weyl fermion, respectively.  The Baby $\otimes$ MW fermionic CFT is invariant under stacking of the (1+1)$d$ invertible spin TQFT $(-1)^{\text{Arf}}$, since all its partition functions with odd spin structure vanish. Correspondingly, the Monster CFT is self-dual under the $\bZ_{2A}$  orbifold.}\label{fig:2A2B}
\end{figure}
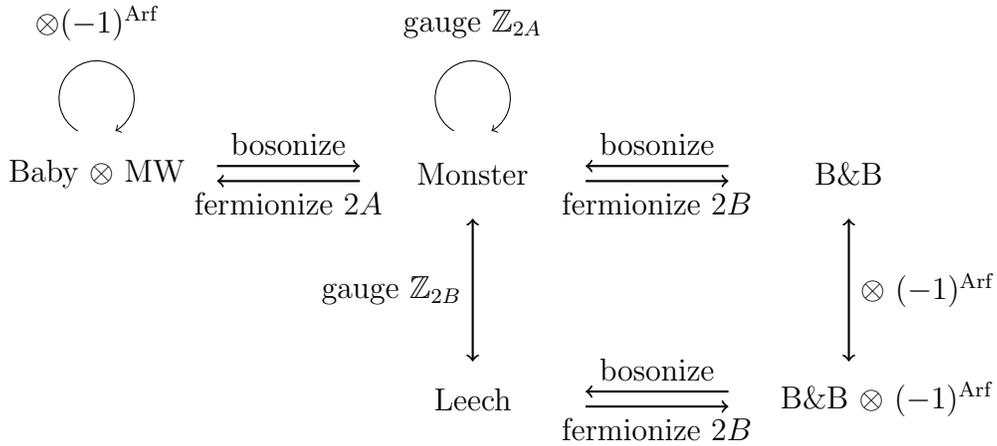

Given any (1+1)$d$ bosonic CFT with a non-anomalous $\bZ_2$ global symmetry, one can fermionize it to obtain a fermionic CFT \cite{Gaiotto:2015zta,Kapustin:2017jrc,Karch:2019lnn,yujitasi,Ji:2019ugf}.  
We find that the fermionization of the Monster CFT with respect to the $\bZ_{2A}$ is the tensor product of the Baby Monster CFT and a Majorana-Weyl fermion, which are holomorphic, fermionic CFTs of central charge $c={47\over2}$ and $c={1\over2}$, respectively.  
On the other hand, the fermionization with respect to $\bZ_{2B}$ gives the ``Beauty and the Beast" ${\cal N}=1$ SCFT \cite{Dixon:1988qd}.  
See Figure \ref{fig:2A2B}.
This gives a physical interpretation for H\"ohn's construction of the Baby Monster super vertex operator algebra ({VOA}) \cite{hoehn2007selbstduale} (see also \cite{Yamauchi}) and Dixon, Ginsparg, Harvey's construction for the  ``Beauty and the Beast"  {VOA}  \cite{Dixon:1988qd} as fermionization in (1+1)$d$ quantum field theory. 
  
The free fermion in the $2A$-fermionized Monster CFT implies that there is a new (non-symmetry) topological defect in the Monster CFT. Below we give an overview of the importance of the topological defects and their relation to global symmetries.

\subsubsection*{Topological Defects as Generalizations of Global Symmetries}

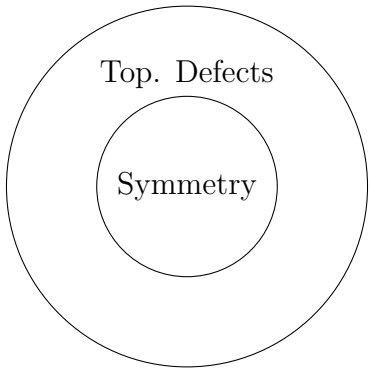
\begin{figure}[h!]
\centering
\begin{tikzpicture}
\draw  (0,0) circle [radius=1.2];
\draw (0,0) node[] {Symmetry};
\draw (0,1.5) node[] {Top. Defects};
\draw  (0,0) circle [radius=2.4];
\end{tikzpicture}
\caption{Topological defects are generalizations of  global symmetries. }\label{fig:ven}
\end{figure}

In quantum field theory, every ($q$-form) global symmetry is associated with a (codimension-$(q+1)$) topological defect (a.k.a. the charge operator) that implements the symmetry action \cite{Kapustin:2014gua,Gaiotto:2014kfa}.  
For a continuous ordinary (0-form) global symmetry, the topological defect is simply the exponentiation of the Noether charge.
The converse is however not true: Not every topological defect is associated with a global symmetry. 
Such a defect is called a non-invertible defect, or sometimes dubbed a non-symmetry defect. 
The definition of a non-invertible defect is that it does not have an inverse under fusion; by contrast, the fusion of symmetry defects obeys the group multiplication rules, which in particular guarantees the existence of an inverse, and hence symmetry defects are invertible.  

It has been advocated that general topological defects should be thought of as generalizations of global symmetries/invertible defects \cite{Bhardwaj:2017xup,Chang:2018iay} (see Figure \ref{fig:ven}).  
In (1+1)$d$, the mathematical framework for the topological defect lines is the theory of {\it tensor categories} \cite{Etingof:aa,etingof2016tensor}, which encodes rules for splitting and joining and should be thought of as generalizations of anomalies.\footnote{If there are only finitely many simple lines in a theory (such as in the minimal models), then the topological defect lines are described by a {\it fusion category}.}
The power of anomaly matching is thereby also generalized.
For instance, it was demonstrated in \cite{Chang:2018iay} that the non-invertible topological defects, just like ordinary global symmetries with 't Hooft anomaly, impose strong constraints on the renormalization group flows in (1+1)$d$. 

In (1+1)$d$,  it has been known for a long time that there exist {\it duality defect lines} \cite{Frohlich:2004ef,Frohlich:2006ch,Frohlich:2009gb,Chang:2018iay,Ji:2019ugf} that are not associated with any ordinary global symmetry. 
The simplest example is the  defect line  $\cal N$ implementing the Kramers-Wannier duality in the $c_L=c_R=\frac 12$ Ising CFT.  
The duality defect line $\cal N$ is not invertible because it obeys the fusion rule ${\cal N} \times {\cal N} = I+\eta$, where $\eta$ is the (invertible) topological defect line for the $\bZ_2$ global symmetry of the Ising CFT.  Hence, the full fusion category of the (1+1)$d$ Ising CFT includes both the $\bZ_2$ symmetry line $\eta$ and the duality defect line $\cal N$.

Therefore, understanding the topological defects is as fundamental as the understanding the global symmetry of a quantum field theory.

\subsubsection*{Duality Defects of the Monster CFT}
 
With the modern perspective that topological defects are generalizations of global symmetry, it is natural to ask:
\textit{What are the topological defect lines in the Monster CFT?}  

It turns out that just like the Ising CFT, the Monster CFT has a Kramers-Wannier duality defect line $\cal N$.\footnote{In the Ising model, the Kramers-Wannier duality refers to the equivalence between the high temperature and low temperature phases away from the critical point. When traced back to the critical point, it reduces to the statement that the Ising CFT is self-dual under the $\bZ_2$ orbifold.  The Monster CFT, by contrast, does not have any relevant deformation, and hence no low temperature phase to speak of.  By the Kramers-Wannier duality of the Monster CFT, we only refer to its self-duality under the $\bZ_{2A}$ orbifold.}    
This is expected by the observation that the Monster CFT is self-dual under the $\bZ_{2A}$ orbifold, but the $\cal N$ line can be identified even more explicitly as follows. The $\bZ_{2A}$ fermionization of the Monster CFT has a chiral fermion parity $\bZ_2^\psi$ with one unit of the mod 8 't Hooft anomaly. 
Under bosonization, it was shown in \cite{Thorngren:2018bhj,Ji:2019ugf} (see also \cite{Bhardwaj:2016clt}) that any fermionic $\bZ_2$ symmetry with this anomaly becomes the duality defect $\cal N$ of an Ising category in the bosonized theory.

We have therefore identified a non-invertible  topological defect line in the Monster CFT.  Since the duality defect $\cal N$ does not commute with the Monster group, it extends the latter into a larger tensor category of topological defect lines that contains the Ising category as a subcategory (see Table \ref{tab:isingMonster}).\footnote{The Ising category of the Monster CFT can be used to couple the CFT to the (2+1)$d$ Ising $\times$ $\overline{\text{Ising}}$ TQFT as follows.  We first couple the Monster CFT to a (2+1)$d$ $\bZ_2$ gauge theory by gauging the $\bZ_{2A}$ global symmetry. Because of the duality defect,  the (1+1)$d$ boundary preserves the 0-form $\bZ_2^{em}$ global symmetry of the (2+1)$d$ $\bZ_2$ gauge theory that exchanges the electric and the magnetic anyons \cite{Ji:2019eqo,Ji:2019ugf}.  We then  gauge the $\bZ_2^{em}$ global symmetry, and the bulk becomes the Ising $\times$ $\overline{\text{Ising}}$ TQFT \cite{Barkeshli:2014cna}.  The partition functions of this 2$d$-3$d$ system with different anyon lines inserted are given by linear combinations of \eqref{Monster10}.}   
This Ising subcategory commutes with the double cover   of the Baby Monster group $2.\mathbb{B}$, which is the centralizer of $\bZ_{2A}$ in the Monster group.

\begin{table}
\ie
\left.
\renewcommand{\arraystretch}{2}
\begin{array}{|c|ccc|}
\hline
& ~~~~\text{Global Symmetry}~~~~~~~~&\subset&~~~~ \text{Topological Defects}~~~~
\\ 
\vspace{-.1in} & \vspace{-.1in} & \vspace{-.1in} & 
\vspace{-.1in} 
\\
\hline
~ \text{Ising CFT}~ & \bZ_2  &\subset& ~~~~\text{Ising Category}=\{I,\eta,{\cal N}\}~~~~ \\
&&&\cap
\\
~~\text{Monster CFT}~~ & \text{Monster Group} & \subset & \text{???}
\\\hline 
\end{array}\right.
\notag
\fe
\caption{The full set of topological defect lines for the Monster CFT is not known, but it contains the (invertible) Monster symmetry lines and the (non-invertible) duality defect line $\cal N$.  The duality defect line $\cal N$ and the $\bZ_{2A}$ line $\eta$ together form an Ising category, which is also realized in the Ising CFT.} \label{tab:isingMonster}
\end{table}

\subsubsection*{Defect McKay-Thompson Series}

The global symmetry of the Monster CFT allows us to  twist the partition function (in the time direction) with different elements of the Monster group.\footnote{In some literature, the partition function with a $g$-twist in the time direction is called the $g$-twined partition function, while a $g$-twist in the space direction is called the $g$-twisted partition function.  We will use the word ``twist" for both the space and time directions. } The resulting partition functions are known as the McKay-Thompson series.   
In the spirit that topological defects are generalizations of global symmetry, we should also consider partition functions twisted by the more general topological defect lines. 
We call the partition function  twisted by non-invertible topological defects in the time direction the {\it defect McKay-Thompson series}. 
We will compute the torus partition function  twisted by the duality defect $\cal N$ using the fermionization of the Monster CFT, and show that its coefficients can be decomposed into the dimensions of the $2.\mathbb{B}$ irreps.  
We will also compute the more general partition functions twisted by both the duality defect line $\cal N$ line and the $\bZ_{2A}$ line $\eta$.  

The usual McKay-Thompson series has the remarkable property that it is the Hauptmodul of a genus-zero subgroup of $PSL(2,\mathbb{R})$ \cite{Conway,Frenkel:1988xz,Borcherds1992} (see \cite{Paquette:2016xoo,Paquette:2017xui} for physics explanations).  
Since the defect McKay-Thompson series is a natural generalization, it is interesting to investigate its modular property.  
We identify a genus-zero congruence subgroup $16D^0$ of $PSL(2,\bZ)$ that leaves the defect McKay-Thompson series invariant.  
However, by examining its poles, in particular the ones at $\tau=0, i\infty$, we show that the defect McKay-Thompson series cannot be the Hauptmodul of its invariance group inside $PSL(2,\mathbb{R})$.

\section{Bosonic and Fermionic CFTs}

It is important to distinguish between bosonic (non-spin) and fermionic (spin) QFTs. The former can be defined on manifolds without a choice of the spin structure, while the latter requires such a choice.  
For example,  the Ising CFT with $c_L=c_R=\frac12$ (a.k.a.\ the $(3,4)$ minimal model) is a bosonic CFT whose torus partition function is unique. By contrast, a Majorana fermion (also with $c_L=c_R=\frac12$) is a fermionic CFT that has four torus partition functions associated with the four spin structures.

The most familiar bosonic CFTs are those that are invariant under the   modular transformations ({\it e.g.} the Ising CFT).  In particular, the modular invariance on a torus  requires that the chiral central charge be a multiple of 24, {\it i.e.} $c_L-c_R\in 24\bZ$.
More generally, we can relax this condition, and consider bosonic CFTs whose modular non-invariance can be cured by a (2+1)$d$ bosonic gravitational Chern-Simons term. Since the latter has chiral central charge a multiple of 8, this more general class of bosonic CFTs must have $c_L-c_R\in 8\bZ$.  The classic example is the  holomorphic $(E_8)_1$ VOA with $c_L=8$ and $c_R=0$.

For a fermionic CFT ({\it e.g.} a Majorana fermion), we do not require modular {\it invariance} because different spin structures are generally permuted under modular transformations. 
Instead, we require that the partition functions associated with different spin structures transform {\it covariantly} under the modular transformations.   
Similar to the bosonic case, we can consider more general fermionic CFTs whose partition functions are modular covariant up to certain phases that can be canceled by a (2+1)$d$ fermionic gravitational Chern-Simons term. Since the latter has chiral central charge a multiple of $\frac12$, we allow the chiral central charge of  a fermionic  CFT to be $c_L-c_R\in {\bZ\over2}$.   
The simplest example is a Majorana-Weyl fermion with $c_L=\frac 12$ and $c_R=0$.

\subsection{Global Symmetry of Fermionic Theories}\label{sec:globalsymmetry}

In any fermionic theory, there is a canonical $\bZ_2$ global symmetry that shifts the spin structure.  We will call it the fermion parity $(-1)^F$.  
There are two sectors of operators in any fermionic theory, the NS sector and the R sector. 
The NS sector operators are local operators with integer or half-integer spin $s$, and the fermion parity $(-1)^F$ acts as $(-1)^{2s}$ on them.  By contrast, the R sector operators have spins that can be multiples of $1\over16$, and can be thought of as non-local operators attached to the $(-1)^F$ defect line.  
 
The global symmetry of a fermionic theory is defined to act on the NS sector local operators by linear representations.  
By contrast, on the R sector operators, the global symmetry might act projectively, or even become a non-group action \cite{Thorngren:2018bhj,Ji:2019ugf}.  
For example, consider $N$ Majorana-Weyl fermions.  
The global symmetry contains an $SO(N)$, which   rotates the $N$ chiral fermions (which are in the NS sector) in the vector representation.  
The R sector operators are in the spinor representations of $Spin(N)$, {\it i.e.} they are in the projective representations of the global symmetry $SO(N)$. 

Finally, the 't Hooft anomaly of a $\bZ_2$ global symmetry in a fermionic theory has a $\bZ_8$ classification  \cite{Ryu:2012he,Yao:2012dhg,Gu:2013azn,Kapustin:2014dxa}.  For example,  the $(-1)^F$ fermion parity of  $N$ Majorana-Weyl fermions has $N$ mod 8 anomaly.


%

\subsection{Torus Partition Functions of Fermionic CFTs}

On the torus, there are four spin structures whose corresponding partition functions sometimes admit interpretations as traces over Hilbert spaces:
\ie
&Z_{NS} (\tau,\bar \tau ) = \Tr_{{\cal H}_{NS}} [q^{h-c_L/24}\bar q^{\bar h-c_R/24} ]  \,,\\
&\tilde Z_{NS} (\tau,\bar \tau ) = \Tr_{{\cal H}_{NS}} [\, (-1)^F\, q^{h-c_L/24}\bar q^{\bar h-c_R/24 }]  \,,\\
&Z_{R} (\tau,\bar \tau ) = \Tr_{{\cal H}_{R}} [q^{h-c_L/24} \bar q^{\bar h-c_R/24}]   \,,\\
&\tilde Z_{R} (\tau,\bar \tau ) = \Tr_{{\cal H}_{R}} [\, (-1)^F\, q^{h-c_L/24}\bar q^{\bar h-c_R/24} ]  \,.
\fe
Under the modular transformations, the four partition functions transform covariantly as
\ie\label{T}
T:~~&Z_{NS}(\tau+1,\bar \tau+1) = e^{- 2\pi i {c_L-c_R\over 24}} \tilde Z_{NS}(\tau,\bar \tau)\,,
\\
& \tilde Z_{NS}(\tau+1,\bar \tau+1) =e^{- 2\pi i {c_L-c_R\over 24}} Z_{NS}(\tau,\bar \tau)\,,\\
& Z_R(\tau+1,\bar \tau+1)  = e^{2\pi i {c_L-c_R\over 12}} Z_{R}(\tau,\bar \tau)\,,
\\
& \tilde Z_R(\tau+1,\bar \tau+1)  = e^{2\pi i {c_L-c_R\over 12}}\tilde  Z_{R}(\tau,\bar \tau) \,;
\\
~
\\
S:~~&Z_{NS}(-1/\tau,-1/\bar \tau) =  Z_{NS}(\tau,\bar \tau)\,,
\\
& \tilde Z_{NS}(-1/\tau,-1/\bar \tau) = Z_{R}(\tau,\bar \tau)\,,\\
& Z_R(-1/\tau,-1/\bar \tau)  =\tilde Z_{NS}(\tau,\bar \tau)\,,
\\
& \tilde Z_R(-1/\tau,-1/\bar \tau)  =\tilde  Z_{R}(\tau,\bar \tau) \,.
\fe
For example, the torus partition functions of a Majorana-Weyl fermion with $c_L={1\over2}$ and $c_R=0$ are
\ie
&Z_{NS} (\tau ) =  \sqrt{\theta_3(\tau)\over \eta(\tau)} \,,~~~~
\tilde Z_{NS} (\tau ) = \sqrt{\theta_4(\tau)\over \eta(\tau)} \,,~~~~
Z_{R} (\tau )  = \sqrt{\theta_2(\tau)\over \eta(\tau)} \,,~~~~
\tilde Z_{R} (\tau )  = 0\,.
\fe
In this case, there is no Hilbert space interpretation for these partition functions because, for example, $Z_R(\tau ) = \sqrt{2} q^{{1\over 16}  -  {1\over 48} }+\cdots$ does not have positive integer degeneracies.  Nonetheless, these partition functions transform covariantly according to \eqref{T} under modular transformations.

\subsection{Bosonization and Fermionization}
\label{sec:bfzation}

As discussed above, the Ising CFT is a bosonic CFT, while a Majorana fermion is a fermionic CFT, and they should be not equated at face value.  
More precisely, they are related by bosonization and fermionization. 
We emphasize that, contrary to the common usage of these terms, bosonization/fermionization stands for a map between (inequivalent) bosonic and fermionic theories, instead of the equivalence of them.

Bosonization and fermionization in (1+1)$d$ have been revisited from a modern point of view in \cite{Gaiotto:2015zta,Kapustin:2017jrc,Karch:2019lnn,yujitasi,Ji:2019ugf} (see also \cite{Bhardwaj:2016clt,Thorngren:2018bhj,Gaiotto:2018ypj,Hsin:2019gvb}). We refer the readers to, for example, Section 3 of \cite{Ji:2019ugf} for a detailed exposition.

Let us start with a fermionic CFT $\cal F$ with $c_L-c_R\in 8\bZ$.\footnote{When $c_L-c_R$ is not a multiple of 8, there is no candidate bosonic theory with such gravitational anomaly, and summing over the spin structures is either inconsistent or gives a fermionic theory.  We thank Kantaro Ohmori and Nathan Seiberg for discussions on this point.} 
We can sum over the spin structures to define the partition function of a bosonic CFT $\cal B$.\footnote{When summing over the spin structures, there is the freedom of putting a minus sign on all the odd spin structure partition functions.  This freedom can be equivalently described by stacking a nontrivial, invertible (1+1)$d$ spin TQFT $(-1)^{\text{Arf}[\rho]}$ with $\cal F$ before the sum.}  
This is the familiar bosonization procedure (also known as the GSO projection \cite{Gliozzi:1976qd}).  
Let $Z_{\cal F}[\rho]$ be the partition function of the fermionic theory on a genus $g$ Riemann surface $\Sigma_g$ with spin structure $\rho$, then the resulting bosonic partition function is
\ie\label{bosonization}
Z_{\cal B}[T]  =  {1\over 2^g} \sum_{s\in H^1(\Sigma_g,\bZ_2)}  Z_{\cal F}[s+\rho]  \, \exp\left[ i\pi  \left(
\int s\cup T  + {\rm Arf}[T+\rho] +{\rm Arf}[\rho]\right)\right]\,,
\fe
where the sum $s\in H^1(\Sigma_g, \bZ_2)$ is over all discrete $\bZ_2$ gauge fields for the $(-1)^F$ symmetry.  Here $T\in H^1(\Sigma_g, \bZ_2)$ is the background $\bZ_2$ gauge field for the emergent $\bZ_2$ global symmetry of the bosonic theory $\cal B$ \cite{Gaiotto:2015zta}.    The Arf invariant $\text{Arf}[\rho]$ is 0 if $\rho$ is even and 1 if $\rho$ is odd.

Conversely, 
we can  fermionize $\cal B$ with respect to the emergent $\bZ_2$ global symmetry to get back to the original $\cal F$.  
The fermionization can be described as first coupling $\cal B$ to the nontrivial, invertible (1+1)$d$ spin TQFT  $(-1)^{\text{Arf}[\rho]}$, and then gauging a diagonal $\bZ_2$ global symmetry:
 \ie\label{fermionization}
Z_{\cal F}[S+\rho ] = {1\over 2^g} \sum_{t\in H^1(\Sigma_g,\bZ_2)} Z_{\cal B}[t] \, \exp\left[
i \pi
\left(
 {\rm Arf}[t+\rho] + {\rm Arf}[\rho] +\int t\cup S
 \right)
 \right]\,,
\fe
where $S\in H^1(\Sigma_g,\bZ_2)$ is the $\bZ_2$ background gauge field for the fermion parity $(-1)^F$ of $\cal F$. 

More generally, if we start with a bosonic CFT, we can fermionize it with respect to {\it any} non-anomalous $\bZ_2$ global symmetry to obtain a fermionic CFT.\footnote{When the $\bZ_2$ is anomalous, the partition function does not only depend on the cohomology class of the discrete $\bZ_2$ gauge field, but also on the explicit representative. This ruins the fermionization procedure  \cite{Ji:2019ugf}.}
 Hence, the fermionization of a bosonic CFT  may not exist or may not be unique; it depends on the choice of a non-anomalous $\bZ_2$ global symmetry.

For two theories related by bosonization/fermionization, what is the relation between their states/operators?
For a bosonic CFT with a non-anomalous $\bZ_2$ global symmetry, we have the ordinary Hilbert space $\cal H$ corresponding to the local operators, and a defect Hilbert space ${\cal H}_{\eta}$, corresponding to non-local, point-like operators living at the end of the $\bZ_2$ topological defect line $\eta$.  
In each of $\cal H$ and ${\cal H}_{\eta}$, we can further divide them into the $\bZ_2$-even/odd subsectors. 
All in all, we end up with four sectors of states in a bosonic  CFT with a non-anomalous $\bZ_2$ symmetry: ${\cal H}^\pm$ and ${\cal H}_\eta^\pm$.\footnote{The $\bZ_2$-even subsector ${\cal H}_\eta^+$ of the defect Hilbert space  becomes the twisted sector in the (bosonic) orbifold theory ${\cal B}/\bZ_2$.}
On the other hand, a general fermionic CFT also has four sectors of states: the $(-1)^F$-even/odd NS sectors ${\cal H}_{NS}^\pm$ and the $(-1)^F$-even/odd R sectors ${\cal H}_{R}^\pm$.  The identification is explained in Table~\ref{BF}.

\begin{table}[H]
\ie\nonumber
\left.
\renewcommand{\arraystretch}{1.75}
\begin{array}{|cc|c|}
\hline ~~{\cal B}~~~~ &~~~ {\cal F}~~~~ & ~~\text{Ising/Maj} ~~\\\hline
~~~~ {\cal H}^+ ~~ =& {\cal H}_{NS}^+ & 1_{0,0}, ~\varepsilon_{\frac 12,\frac12} \\

 ~~~~{\cal H}^- ~~= &{\cal H}_R^+ &  \sigma_{{1\over 16},{1\over16}} \\
  ~~~~{\cal H}_\eta^+ ~~= & {\cal H}_R^- &  \mu_{{1\over 16},{1\over16}}\\
   ~~~~{\cal H}_\eta^- ~~= & {\cal H}_{NS}^- & \psi_{\frac 12,0}, ~ \overline\psi_{0,\frac 12}
 \\\hline \end{array}
 \right.
\fe
\caption{Identification of Hilbert spaces between a pair of bosonic CFT $\cal B$ and fermionic CFT $\cal F$ related by bosonization/fermionization. In the last column we list the Virasoro primaries in each of the four sectors in the case of ${\cal B} = $ Ising CFT and ${\cal F} = $ Majorana fermion.  The subscripts of an operator ${\cal O}_{h,\bar h}$ denote its conformal weights $h,\bar h$.}
\label{BF}
\end{table}

\subsection{Baby Monster as a Fermionic CFT}\label{sec:babyfermion}

The Baby Monster VOA \cite{hoehn2007selbstduale} is a VOA with $c = {47 \over 2}$ that has three irreducible modules $V_h^{\rm Baby}$ labeled by their conformal weight $h=0, \, {3\over2}, \, {31\over16}$.  
We interpret the $h=0, \, {3\over2}$ modules as the NS sector operators (together they are known mathematically  as the Shorter Moonshine Module), and the $h={31\over 16}$ operators as the R sector operators. 
Some more details of the construction as well as the characters and their modular properties are given in Appendix~\ref{App:VOA}.

The Baby Monster {VOA} is realized in a fermionic CFT with $c_L= {47\over2}$ and $c_R=0$, whose torus partition functions are
\ie\label{baby4}
&Z_{NS}^{\rm Baby}(\tau) = \chi^{\rm Baby} _0 (\tau) + \chi^{\rm Baby}_{3\over2}(\tau) \,,
\\
& \tilde Z_{NS}^{\rm Baby}(\tau)  = \chi^{\rm Baby} _0 (\tau) - \chi^{\rm Baby}_{3\over2}(\tau) \,,
\\
&Z_R^{\rm Baby}(\tau) = \sqrt{2} \chi^{\rm Baby}_{31\over16}(\tau) \,,
\\
& \tilde Z_R^{\rm Baby}(\tau)  = 0 \,.
\fe
Using the modular $S$ and $T$ matrices \eqref{TS},  one can check that these four partition functions obey the modular covariance of a fermionic CFT \eqref{T}. 
Just like a single Majorana-Weyl fermion, the Ramond sector partition function $Z^{\rm Baby}_R$ does not admit a Hilbert space interpretation.  

The global symmetry of the Baby Monster CFT  is $(-1)^F \times \mathbb{B}$ \cite{Hoehn}, where $\mathbb{B}$ is the Baby Monster group, the second largest sporadic finite simple group.    
As discussed in Section \ref{sec:globalsymmetry}, while the global symmetry of a fermionic theory by definition acts linearly on the NS sector, it might act projectively on the R sector in the presence of a mixed anomaly between $(-1)^F$ and the global symmetry.  
It is known from Lemma 4.2.10 of \cite{hoehn2007selbstduale} that the modules $V^{\rm Baby}_0$ and $V^{\rm Baby}_{3\over2}$ can each be decomposed into irreps of $\mathbb{B}$. 
On the other hand, $V^{\rm Baby}_{31\over16}$ can only be decomposed into {\it projective} irreps of $\mathbb{B}$, {\it i.e.} they are the irreps of the double cover $2.\mathbb{B}$ of the Baby Monster group, but not of $\mathbb{B}$. 
This is the effect of the mixed anomaly between $(-1)^F$ and $\mathbb{B}$ in the Baby Monster CFT that extends $\mathbb{B}$ to its double cover in the R sector.

The consistency of this theory beyond the torus partition functions has been studied in~\cite{Mukhi:2017ugw}.  In particular, they computed the four-point conformal blocks for the Baby Monster VOA, but due to the large degeneracies of Baby Monster primaries, the OPE data cannot be simply solved by imposing crossing symmetry.  In the next section, we provide an explicit construction of the Baby Monster CFT via fermionization of the Monster CFT.

\section{Monster CFT and its $2A$ Fermionization}
\label{Sec:2AFermionization}

The torus partition function of the (holomorphic) $c_L=24$ and $c_R=0$ Monster CFT is
\ie\label{ZJ}
Z(\tau) = J(\tau )  ={1\over q}  + 196884q+   21493760q^2+\cdots,
\fe
where $J = j -744$ with $j$ being Klein's $j$-invariant. 
The Monster CFT has two non-anomalous $\bZ_2$ symmetries, usually denoted by $\bZ_{2A}$ and $\bZ_{2B}$.  
We simply note that the fermionization with respect to the $\bZ_{2B}$ gives the ``Beauty and the Beast" ${\cal N}=1$ SCFT, and defer some more details to Appendix~\ref{App:BB}.
We now discuss in detail the fermionization with respect to the $\bZ_{2A}$.  

We will denote the topological defect line associated with the $\bZ_{2A}$ symmetry as $\eta$. 
The torus partition function $Z^{\eta}$ with a topological line $\eta$ {extended} along the space direction  ({\it i.e.} a $\bZ_{2A}$ twist in the time direction, see \eqref{Monster10}) is
\ie\label{Zetat}
Z^{\eta}  (\tau) &=\Tr_{\cal H} \left[\hat\eta \, q^{h-c_L/24} \right]
=
  {\eta(\tau )^{24}\over\eta(2\tau)^{24} } +2^{12} {\eta(2\tau )^{24}\over\eta(\tau)^{24} }  +24
\\
&=  {1\over q}  +4372 q+ 96256 q^2 +1240002q^3+{\cal O}(q^4)\,,
\fe
where $\hat\eta:{\cal H}\to {\cal H}$ represents the $\bZ_{2A}$ charge operator obeying $\hat \eta \circ \hat \eta=1$.\footnote{Throughout this paper, the superscript (subscript) of $\eta$ or $\cal N$ for the partition function means a twist  in the time (space) direction, which is equivalent to putting the corresponding topological defect line {extended} along the space (time) direction.}

The modular $S$ transform of \eqref{Zetat} is the partition function  $Z_{\eta}$  with  a topological line $\eta$ {extended} along the time direction  ({\it i.e.} a $\bZ_{2A}$ twist in the space direction) \cite{Gaiotto:2008jt},
\begin{align}\label{Zetax}
Z_{\eta} (\tau) &=\Tr_{{\cal H}_\eta} \left[q^{h-c_L/24} \right]
= 
2 ^{12} {\eta(\tau)^{24} \over \eta(\tau/2) ^{24}  }  +{\eta(\tau/2) ^{24}\over \eta(\tau)^{24}} +24
\\
&= {1\over q^{1/2} }
+4372 q^{1/2}  + 96256  q  + 1240002 q^{3/2} +{\cal O}(q^2)\,,\notag
\end{align}
where ${\cal H}_\eta$ is the defect Hilbert space obtained by quantizing the Monster CFT with a $\bZ_{2A}$ twist in the space direction. 
The torus partition function with twists in both the time and space directions is obtained by applying a $T$ transform on $Z_{\eta}(\tau)$, {\it i.e.} $Z_\eta^\eta(\tau) = Z_\eta(\tau+1)$. 
 
Let us consider the fermionization of the bosonic Monster CFT with respect to the $\bZ_{2A}$ symmetry.  We will denote this fermionic theory by $\cal F$.  
Its NS sector torus partition function is given by (see \eqref{fermionization} and Table~\ref{BF} for the general fermionization prescription)
\ie
\label{ZNSF}
Z_{NS}^{\cal F}(\tau ) = \frac12 \left[ \, 
Z(\tau)  +  Z^{\eta}(\tau ) + Z_{\eta}(\tau) -  Z_{\eta}^\eta(\tau)
\,\right]
= {1\over q}  +{1\over \sqrt{q} }  +4372 \sqrt{q}  +100628 q+\cdots\,.
\fe
Similarly, the R sector torus partition function is
\ie
Z_R^{\cal F}(\tau )  =\frac 12\left[\,
Z(\tau)  -  Z^{\eta}(\tau ) + Z_{\eta}(\tau) +Z_{\eta}^\eta(\tau)
\,\right]
=192512q + 21397504 q^2 + 863059968 q^3+\cdots\,.
\fe
The signs in the above formulae are determined by the Arf invariant and the cup product between the $\bZ_2$ gauge fields as in \eqref{fermionization}.  

The presence of the $1\over\sqrt{q}$ term in $Z_{NS}^{\cal F}$ means that $\cal F$ contains a single free Majorana-Weyl fermion $\psi(z)$ of weight $h={1\over2}$.  
This means that $\cal F$ takes the form\footnote{It was proven in \cite{Goddard:1988wv} that a holomorphic CFT can always be decomposed into the tensor product of a sector of free Majorana-Weyl fermions (which might be empty) and a sector without any $h={1\over2}$ operator.}
\ie
{\cal F} = {\cal F}' \otimes \text{MW fermion}\,,
\fe
where ${\cal F}'$ is a fermionic CFT with $c_L=  23+\frac 12$ and $c_R=0$ (which will be identified as the Baby Monster CFT momentarily).  
We can thus write the NS sector partition function of $\cal F$ as
\ie\label{NSZ}
Z_{NS}^{\cal F}(\tau )  =Z_{NS}^{\rm Ising}(\tau)  \, Z_{NS}^{{\cal F}'}(\tau)  \,,
\fe
with $Z_{NS}^{\rm Ising} = \sqrt{\theta_3\over\eta}$ and $Z_{NS}^{\cal F}$ given in \eqref{ZNSF}. Inverting this relation, we find that 
the NS sector partition function of ${\cal F}'$
is (see Appendix~\ref{App:VOA} for the characters of the Baby Monster VOA)
\ie
Z_{NS}^{{\cal F}'}(\tau) &  ={1\over q^{47/48}}+4371 q^{25/48}+ 96256 q^{49/48}+1143745 q^{73/48}+\cdots\\
&= \chi^{\rm Baby} _0 (\tau)+ \chi^{\rm Baby}_{3/2}(\tau)\,.
\fe
This is precisely the Baby Monster CFT partition function \eqref{baby4}.
Thus, we identify ${\cal F}'$ as the Baby Monster CFT.
  Indeed, the R sector partition function of $\cal F$ can also be written as
\ie
Z_R^{\cal F}(\tau ) = Z_R^{\rm Ising}(\tau) \,Z_R^{\rm Baby}(\tau)   
= \sqrt{\theta_2(\tau)\over \eta(\tau)} \times \sqrt{2}\chi_{31\over16}^{\rm Baby}(\tau)\,,
\fe
where the R sector partition function of the Baby Monster CFT is $Z_R^{\rm Baby}= \sqrt{2}\chi_{31\over16}^{\rm Baby}$.  

From \eqref{fermionization}, we can further compute the NS and R sector partition functions of $\cal F$ with $(-1)^F$ inserted: 
\ie\label{-1F}
\tilde Z_{NS}^{\cal F}(\tau ) &= \frac12 \left[ \, 
Z(\tau)  +  Z^{\eta}(\tau ) - Z_{\eta}(\tau) +Z_{\eta}^\eta(\tau)
\,\right]
= {1\over q} -{1\over \sqrt{q} }  -4372 \sqrt{q}  +100628 q+\cdots\,,
\\
\tilde Z_R^{\cal F}(\tau ) &= \frac 12\left[\,
Z(\tau)  -  Z^{\eta}(\tau ) - Z_{\eta}(\tau) -Z_{\eta}^\eta(\tau)
\,\right]
=0\,.
\fe
Note that $\tilde Z_R=0$ because of the zero mode from the Majorana-Weyl fermion in $\cal F$.  Again, we find that the partition functions of $\cal F$ factorize into the product of that of the Baby Monster CFT and the free Majorana-Weyl fermion, $\tilde Z_{NS}^{\cal F} (\tau)  = \tilde Z_{NS}^{\rm Ising}(\tau) \,\tilde Z_{NS}^{\rm Baby}(\tau)  $, $\tilde Z_{R}^{\cal F} (\tau)  = \tilde Z_{R}^{\rm Ising}(\tau) \,\tilde Z_{R}^{\rm Baby}(\tau)$ (of course, the latter is trivially true).  

In \cite{hoehn2007selbstduale,Yamauchi}, it was shown that the $\bZ_{2A}$-invariant ${\cal H}^+$ sector of the Monster VOA is the ${\cal H}_{NS}^+$ sector of a free Majorana-Weyl fermion times an SVOA. The latter was given the name Baby Monster SVOA since  it has the Baby Monster symmetry. 
This in particular implies that the partition function for ${\cal H}^+$ in the Monster CFT equals that for ${\cal H}^+_{NS}$ of a free Majorana-Weyl fermion times the Baby Monster CFT. 
In the above, we further check that the partition functions for the other three sectors ${\cal H}^-, \, {\cal H}^+_\eta, \, {\cal H}^-_\eta$ of the Monster agree with the ${\cal H}_R^+, \, {\cal H}_{R}^- , \, {\cal H}_{NS}^-$ of a free Majorana-Weyl fermion times the Baby Monster CFT.

To conclude, we have provided a new physical interpretation of the construction of the Baby Monster CFT in \cite{hoehn2007selbstduale,Yamauchi} in terms of the fermionization of the Monster CFT with respect to the $\bZ_{2A}$ symmetry:
\ie\label{main}
\nonumber
\boxed{
\text{$2A$ Fermionization of Monster}  = \text{Baby Monster} \otimes \text{MW Fermion}.}
\fe
Conversely, the bosonization ({\it i.e.} summing over the spin structures) of the tensor product of the Baby Monster CFT and the Majorana-Weyl fermion is the Monster CFT.

There is a mathematical identity expressing the $J$-function in terms of  the inner product of the three Ising   and the Baby Monster characters \cite{hoehn2007selbstduale,Yamauchi,Hampapura:2016mmz}: 
\ie\label{bilinear}
J(\tau)=  \chi_0^{\rm Ising}(\tau) \chi_0^{\rm Baby} (\tau)+ \chi_{1\over2}^{\rm Ising}(\tau) \chi_{3\over2}^{\rm Baby} (\tau)  +  \chi_{1\over16}^{\rm Ising} (\tau)\chi_{31\over16}^{\rm Baby}(\tau)\,.
\fe
We can rewrite the above in terms of the torus partition functions of the four spin structures:
\ie
J(\tau )  =\frac12 \left[ Z_{NS}^{\rm Ising}(\tau)Z_{NS}^{\rm Baby}(\tau)
+\tilde Z_{NS}^{\rm Ising}(\tau)  \tilde  Z_{NS}^{\rm Baby}(\tau)
+Z_{R}^{\rm Ising}(\tau)Z_{R}^{\rm Baby}(\tau)
+\tilde  Z_{R}^{\rm Ising}(\tau)\tilde  Z_{R}^{\rm Baby}(\tau)\right]\,.
\fe
This is simply the statement that the bosonization  \eqref{bosonization} of the fermionic CFT $\cal F$ (=Baby Monster  $\otimes$   MW Fermion) is the Monster CFT.  
We have therefore given a physical interpretation of the bilinear relation \eqref{bilinear} as  a GSO projection. 

Finally, let us comment what happens to the global symmetry upon bosonization.
As discussed in Section \ref{sec:babyfermion}, the global symmetry of the Baby Monster CFT is $(-1)^F \times \mathbb{B}$. 
Due to the mixed anomaly between $(-1)^F$ and $\mathbb{B}$, the bosonization of $\cal F$ extends the Baby Monster group symmetry $\mathbb{B}$  to its double cover $2.\mathbb{B}$ in the resulting bosonic theory, the Monster CFT. Indeed, the Monster group contains $2.\mathbb{B}$ as a subgroup, but not $\mathbb{B}$.

\section{Duality Defect of the Monster CFT}

The free fermion sector of $\cal F$ has a fermionic $\bZ_2^\psi$ symmetry, the chiral fermion parity, that flips the sign of $\psi(z)$,
\ie
\bZ_2^\psi:~ \psi(z) \mapsto - \psi(z)\,.
\fe
As discussed in Section \ref{sec:globalsymmetry}, this chiral fermion parity has one unit of the mod 8 anomaly.  
Hence, under bosonization, it becomes a duality defect $\cal N$ of the Monster CFT \cite{Thorngren:2018bhj,Ji:2019ugf}.  
See Table~\ref{tab:relation} for the relations between $\bZ_2^\psi$, $(-1)^F$ of $\cal F$ and ${\cal N}$, $\bZ_{2A}$ of the Monster CFT.

The non-invertible duality defect line $\cal N$ is a not a symmetry defect. Rather, it satisfies the fusion rule:
\ie\label{fusion}
{\cal N}\times {\cal N}  = I+\eta\,,~~~\eta\times \eta = I\,,~~~{\cal N}\times\eta =  \eta \times {\cal N} = {\cal N}\,,
\fe
where $\eta$ is the $\bZ_{2A}$ line.  
Indeed, we see that there is no inverse of $\cal N$, so it is not an invertible defect. 
This is an example of how a symmetry defect of a fermionic theory becomes a non-invertible (non-symmetry) defect in the bosonized theory. 
Together with the identity line, they form the Ising category.\footnote{To be more precise, there are two fusion categories obeying the fusion rules \eqref{fusion}, but with different $F$-symbols. They are collectively called the $\bZ_2$ Tambara-Yamagami fusion categories \cite{TAMBARA1998692}.  One of them is realized in the Ising CFT (and will be called the Ising category), while the other is realized in the $SU(2)_2$ WZW model (and will be called the $SU(2)_2$ category).  More generally, the $Spin(N)_1$ WZW model has the Ising category if $N=1,7$ mod 8, and has the $SU(2)_2$ fusion category if $N=3,5$ mod 8. The $F$-symbols of the these categories can be found in \cite{TAMBARA1998692} (see also \cite{Chang:2018iay}). }

It was shown in Section 4.3.1 of \cite{Chang:2018iay} that the existence of the duality defect $\cal N$ implies that the bosonic theory is self-dual under the $\bZ_{2A}$ orbifold 
\ie
 {\text{\rm Monster}\over \bZ_{2A}} =\text{\rm Monster}\,.
\fe      
Indeed, the self-duality of the Monster CFT under the $\bZ_{2A}$ orbifold can be proven using the methods of {\it e.g.} \cite{Paquette:2016xoo,Paquette:2017xui,2017arXiv170702954C}.
As a consistency check, the torus partition function of the $\bZ_{2A}$ orbifold partition function is
\ie\label{selfdual}
\frac 12 \left[ Z(\tau ) +  Z^\eta(\tau ) + Z_\eta(\tau) +Z_\eta^\eta(\tau)\right]
\fe
which equals the original Monster partition function $Z(\tau)=J(\tau)$.

\begin{table}[H]
\begin{align*}
\left.
\renewcommand{\arraystretch}{2.5}
\begin{array}{|ccc|}
\hline
{\cal F}=\text{MW $\otimes$ Baby} & \xleftrightarrow[\text{\small \qquad fermionization \qquad}]{\text{\small \qquad bosonization \qquad}} & {\cal B}=\text{Monster CFT} 
\\
\vspace{-.15in} & \vspace{-.15in} & \vspace{-.15in}
\\
\hline
\text{fermion parity}~~(-1)^F& \xleftrightarrow{\text{\small \quad dual \quad}}  & ~~ \text{non-anomalous}~~\bZ_{2A} ~~
\\
~~ \text{chiral fermion parity} ~~ \bZ_2^{\psi} ~~ & ~~~~ \xrightarrow{\text{\small \quad non-symmetry extension \quad}} ~~~~& \text{duality defect}~~{\cal N}  
\\
\text{Baby Monster group}~~ \mathbb{B}  &\xrightarrow{\text{\small \qquad symmetry extension \qquad}} &\text{double cover}~~2.\mathbb{B}
\\
\hline
\end{array}
\right.
\end{align*}
\caption{Under bosonization/fermionization, the  fermion parity $(-1)^F$  is the emergent quantum symmetry of $\bZ_{2A}$, while the chiral fermion parity $\bZ_2^\psi$ of the Majorana-Weyl fermion is extended to a non-invertible duality defect $\cal N$ in the Monster CFT ${\cal B}$.  The Baby Monster group $\mathbb{B}$ is extended to its double cover due to the mixed anomaly with $(-1)^F$.}\label{tab:relation}
\end{table}

\subsection{Duality Defect of the Ising CFT}

\begin{figure}[H]
\ie\nonumber
\begin{gathered}
\begin{tikzpicture}[scale = 1]
\tikzset{line/.style={line width=0.25mm}}
\draw [line] (0,0) circle (1);
\filldraw [line] (0,0) circle (.05);
\node at (0,-.5) {$\phi(x)$};
\node at (2,0) {$\to$};
\filldraw [line] (4,0) circle (.05);
\node at (4,-.5) {$\widehat{\cal N} \cdot \phi(x)$};
\end{tikzpicture}
\end{gathered}
\fe
\caption{The duality defect line $\cal N$ defines a (non-invertible) map $\widehat{\cal N}:{\cal H}\rightarrow {\cal H}$ on the Hilbert space of local operators  by encircling a local operator $\phi(x)$ and shrinking the circle.}\label{fig:hatN}
\end{figure}
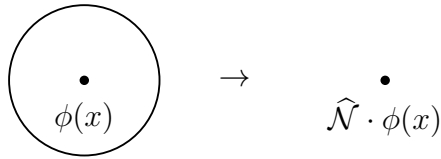

How does the duality defect line act on the local operators? We refer the readers to \cite{Frohlich:2004ef,Chang:2018iay,Ji:2019ugf} for detailed discussions on the duality defect line.  Below we only give a lightening review. 
Using the duality defect $\cal N$, we define a map on a local operator $\phi(x)$ as follows: We circle $\cal N$ around $\phi(x)$, and shrink the circle to produce another local operator (see Figure \ref{fig:hatN}).  We denote this map on the Hilbert space of local operators $\cal H$  as $\widehat{\cal N} :~{\cal H} \rightarrow {\cal H}$. 
Since $\cal N$ is a topological defect line, it commutes with the stress tensor, and hence $\phi$ and $\widehat{\cal N}\cdot \phi$ have the same conformal weights $h, \bar h$.

If $\widehat{\cal N}$ were a symmetry defect, then $\widehat{\cal N}$ would have been the symmetry transformation on local operators and is in particular invertible. 
One characteristic feature of the duality defect is that $\widehat{\cal N}$ is non-invertible. 
From the fusion rule ${\cal N}\times {\cal N}=  I+\eta$, we see that the eigenvalues of $\widehat{\cal N}$ are
\ie
\widehat{\cal N}\cdot \phi= \begin{cases}
&\pm\sqrt{2}~ \phi \,,~~~\text{if $\phi$ is $\bZ_{2A}$-even}\,,\\
&0  \,,~~~~~~~~~~~\text{if}\text{ $\phi$ is $\bZ_{2A}$-odd}\,.
\end{cases}
\fe
Indeed, the map $\widehat{\cal N}$ on the $\bZ_{2A}$-odd operators are non-invertible.   
In particular, an empty loop of the duality defect $\cal N$ is $\sqrt{2}$, which is its quantum dimension. 

Let us illustrate this in the Ising CFT.   The $\bZ_2$ line $\eta$ acts on three primaries as
\ie
\widehat{\eta} \cdot 1= 1\,,~~~\widehat{\eta} \cdot \varepsilon_{{1\over 2} ,{1\over 2}} = \varepsilon_{{1\over 2} ,{1\over 2}}\,,~~~\widehat{\eta} \cdot \sigma_{{1\over 16} ,{1\over 16}} = -\sigma_{{1\over 16} ,{1\over 16}}\,,
\fe
and the duality defect line acts as
\ie
\widehat{\cal N} \cdot 1 = \sqrt{2}\,,~~~\widehat{\cal N} \cdot \varepsilon_{{1\over 2} ,{1\over 2}} = -\sqrt{2}\varepsilon_{{1\over 2} ,{1\over 2}}\,,~~~\widehat{\cal N} \cdot \sigma_{{1\over 16} ,{1\over 16}} = 0\,.
\fe
It follows that the torus partition function with an $\cal N$ line extended along the space direction is
\ie
(\text{Ising}):~Z^{\cal N} (\tau ,\bar \tau   )  =  \text{Tr}_{\cal H} \left[  \widehat{\cal N}\, q^{h-c_L/24} \bar q^{\bar h-c_R/24} \right] =
  \sqrt{2} |\chi_0^{\rm Ising} (\tau)|^2-   \sqrt{2} |\chi_{1\over2}^{\rm Ising} (\tau)|^2\,.
\fe
By the modular $S$ transformation, we immediately get the defect Hilbert space ${\cal H}_{\cal N}$ for the duality defect line:
\begin{align}
(\text{Ising}):~&Z_{\cal N} (\tau, \bar\tau  )  =  \text{Tr}_{ {\cal H}_{\cal N}} \left[ q^{h-c_L/24} \bar q^{\bar h-c_R/24} \right] = Z^{\cal N} (-1/\tau  ,-1/\bar\tau) \\
&=
  \chi_0^{\rm Ising} (\tau)\chi_{1\over16} ^{\rm Ising} (\bar \tau)
+    \chi_{1\over2}^{\rm Ising} (\tau)\chi_{1\over16} ^{\rm Ising} (\bar \tau)
+      \chi_{1\over16}^{\rm Ising} (\tau)\chi_{0} ^{\rm Ising} (\bar \tau)
  +      \chi_{1\over16}^{\rm Ising} (\tau)\chi_{1\over2} ^{\rm Ising} (\bar \tau)\,.\notag
\end{align}
The partition function $Z_{\cal N}$ is realized by inserting a duality defect line along the time direction, and thereby twisting the space direction.  
We see that the defect Hilbert space ${\cal H}_{\cal N}$ contains four primaries with conformal weights $(0,{1\over16}), ({1\over2},{1\over16}), ({1\over16},0), ({1\over16},{1\over2})$. 
These correspond to non-local operators living at the end of the duality defect line $\cal N$, so they do not need to be have integer spins.  
In fact, their spins are constrained to be $s\equiv h-\bar h   \in \pm {1\over 16}  + {\bZ\over 2}$ \cite{Chang:2018iay}.

\subsection{Defect McKay-Thompson Series}

In the following we will compute the torus partition functions of the Monster CFT twisted   by  the duality defect line $\cal N$. 
In the fermionic theory $\cal F$, the NS sector torus partition function with a $\bZ_2^\psi$ line {extended} along the space direction is
\ie
Z_{NS}^{\psi}(\tau)  = \sqrt{\theta_4(\tau) \over\eta(\tau)} Z_{NS}^{\rm Baby}(\tau)
= \sqrt{\theta_4(\tau)\over \theta_3(\tau)}  Z^{\cal F}_{NS}(\tau)
\,.
\fe
Recall that the NS sector of $\cal F$ consists of the $\bZ_{2A}$-even subsector ${\cal H}^+$  and the $\bZ_{2A}$-odd subsector  ${\cal H}_{\eta}^-$. 
From the spin selection rules derived in \cite{Chang:2018iay,Lin:2019kpn}, ${\cal H}^+$ only contains integer spin operators while ${\cal H}_\eta^-$ only contains half-integer spin operators.  Therefore we can single out $Z^{\cal N}$ by selecting the integer spin terms of $Z_{NS}^\psi$:
\ie\label{ZNt}
 Z ^{\cal N}  (\tau )  &=   \text{Tr}_{\cal H} \left[  \widehat{\cal N}\, q^{h-c_L/24}  \right] 
=\sqrt{2} \times \frac 12 \left[ \, Z_{NS}^\psi(\tau)+Z_{NS}^\psi(\tau+1)\,\right]\\
&= \sqrt2 \, \chi_{0}^{\rm Ising} \chi_0^{\rm Baby} - \sqrt2 \, \chi_{1\over2}^{\rm Ising} \chi_{3\over2}^{\rm Baby}
\\
&=  \sqrt{2} \left(\,
{1\over q} +91886 q + 8498776 q^2+301112552 q^3+\cdots
\,\right)
\,,
\fe
where we have written $Z^{\cal N}$ in terms of the characters of the Ising CFT and the Baby Monster CFT. 
The overall factor of $\sqrt{2}$ is included such that the duality defect line $\cal N$ acts on the identity with the correct eigenvalue. 

Recall that the $2A$ McKay-Thompson series $Z^\eta$ is the torus partition function with a $\bZ_{2A}$ twist in the time direction.  
In this spirit, we call $Z^{\cal N}$ the {\it defect McKay-Thompson series} where the $\bZ_{2A}$ symmetry line is replaced by the non-invertible duality defect $\cal N$.  
Unlike its cousin, the $2A$ McKay-Thompson series $Z^\eta$, the defect McKay-Thompson series $Z^{\cal N}$ is not invariant under $\Gamma_0(2)+$.

Next, we perform the modular $S$ transformation on $Z^{\cal N}$ to obtain the spectrum of the defect Hilbert space of the duality line $\cal N$:
\begin{align}\label{ZNx}
Z_{\cal N} (\tau) &  =  \text{Tr}_{ {\cal H}_{\cal N}} \left[ q^{h-c_L/24}  \right]  
= {1\over \sqrt{2}  } \sqrt{\theta_2(\tau )\over\theta_3(\tau)} Z_{NS}^{\cal F}(\tau)  
+ {1\over \sqrt{2}  } \sqrt{\theta_3(\tau )\over\theta_2(\tau)} Z_R^{\cal F}(\tau)\\
& =  \chi_{1\over16}^{\rm Ising}  \, (\chi_0^{\rm Baby} + \chi_{3\over2}^{\rm Baby} ) 
+  (  \chi_{0}^{\rm Ising} +  \chi_{1\over2}^{\rm Ising} ) \,\chi_{31\over16} ^{\rm Baby} \notag\\
&= {1\over q^{15/16} }  + q^{{1\over16}}  + 4371 q^{9/16} + 96256 q^{15/16}+96257 q^{17/16} +  96256 q^{23/16}+\cdots\,.\notag
\end{align}
The conformal weights $h$ are consistent with the spin selection rule for the defect Hilbert space ${\cal H}_{\cal N}$ derived in (6.5) of \cite{Chang:2018iay}:
\ie\label{spinZN}
h  \in  \pm {1\over 16}  + {\bZ\over2}\,.
\fe

\subsection{Modular Properties}
\label{Sec:Mod}

Let us discuss some properties of the defect McKay-Thompson series $Z^{\cal N}(\tau)$.   
Let $\Gamma_0(N)$ be the congruence subgroup of $PSL(2,\mathbb{Z})$ defined as 
\ie
\label{Gamma0}
\Gamma_0(N)  =\left \{ \left(\begin{array}{cc}a& b \\ c& d\end{array}\right) \in PSL(2,\mathbb{Z})  ~\Big|~  c= 0~\text{mod}~N\right\}\,.
\fe
We will show that $Z^{\cal N}(\tau)$ is invariant under $\Gamma_0(8)$ \textit{up to a sign}.   
The generators of $\Gamma_0(8)$ (viewed as a subgroup of $PSL(2,\mathbb{Z})$ instead of $SL(2,\mathbb{Z})$) are 
\ie\label{TGG}
T=\left(\begin{array}{cc}1& 1 \\0& 1\end{array}\right) \,,~~~G_1 = \left(\begin{array}{cc} 5& -1 \\ 16& -3\end{array}\right) \,,~~~~G_2  =\left(\begin{array}{cc}5& -2 \\ 8& -3\end{array}\right) \,.
\fe
Using the expression for $Z^{\cal N}$ in terms of the Ising and the Baby Monster characters \eqref{ZNt}, as well as their modular $S$ and $T$ matrices \eqref{TS}, we find
\ie
Z^{\cal N} (G_1 \cdot \tau )  = - Z^{\cal N}(\tau)\,,~~~~Z^{\cal N} (G_2 \cdot \tau )  =  Z^{\cal N}(\tau)\,,~~~~
\fe
and of course $Z^{\cal N}(\tau +1)  = Z^{\cal N}(\tau)$.  Hence $\Gamma_0(8)$ leaves $Z^{\cal N}$ invariant up to a sign.\footnote{The invariance up to a sign can be seen quite easily for a particular element $ST^8S$ of $\Gamma_0(8)$. From the spin selection rule \eqref{spinZN}  or the explicit expression \eqref{ZNx}, we see that $Z_{\cal N}(T^8 \cdot \tau ) =  - Z_{\cal N}(\tau)$, which then implies 
$ Z^{\cal N}(ST^8S \cdot \tau ) =  - Z^{\cal N}(\tau)$.}

The defect McKay-Thompson series is invariant under the subgroup of $\Gamma_0(8)$ that 
consists of the words formed by $T, G_1,G_2$ with an even number of $G_1$.
In Appendix~\ref{App:16D} we show that this subgroup is  (the name of $16D^0$ is taken from the classification of genus-zero congruence subgroups by \cite{pauli})
\ie\label{16D0}
16D^0 = \left\{ \left(\begin{array}{cc}a& b \\ c& d\end{array}\right) \in PSL(2,\mathbb{Z}) ~\Big|~ 
\begin{array}{c}
(a=d=\pm1 ~\text{mod}~8\,,~~c=0~\text{mod}~16)
\\
\text{or } (a=d=\pm3 ~\text{mod}~8\,,~~c=8~\text{mod}~16)
\end{array}
\right\} \, .
\fe 
Below we record some of its known properties:
\begin{itemize}
\item It is genus-zero.
\item It is a congruence subgroup at level 16, since we have $\Gamma(16) < \Gamma_1(16) < 16D^0 < \Gamma_0(8)$.\footnote{The level $N$ of a congruence subgroup $\Gamma$ is defined as the smallest $N$ such that $\Gamma(N) \le \Gamma$, where
\ie
\Gamma(N)  =\left \{ \left(\begin{array}{cc}a& b \\ c& d\end{array}\right) \in PSL(2,\mathbb{Z})  ~\Big|~  a,d = 1~\text{mod}~N\,,~~b,c= 0~\text{mod}~N\right\}\,.
\fe
}
\item It is an index 24 subgroup of $PSL(2,\bZ)$.
\item Its cusps and their widths are:
\ie
\left.
\renewcommand{\arraystretch}{2}
\begin{array}{|c|c|c|c|c|c|c|}
\hline \text{cusp} & i\infty & ~0~ & \displaystyle ~{1\over2}~ & \displaystyle ~{1\over4}~ & \displaystyle ~{1\over 6}~ & \displaystyle ~{1\over 8}~ \\\hline \text{width} & 1 & 16 & 2 & 2 & 2 & 1 \\\hline \end{array}
\right.
\fe
\end{itemize}
Unlike the usual McKay-Thompson series, $Z^{\cal N}$ does not appear to be invariant under $\Gamma_0(N)$ for any $N$. In particular, $16D^0$ does not contain any $\Gamma_0(N)$ as a subgroup.

Let $\Gamma_{\cal N}$ be the invariance group of $Z^{\cal N}$ inside $PSL(2,\mathbb{R})$ (instead of $PSL(2,\mathbb{Z})$).  
Could the defect McKay-Thompson series $Z^{\cal N}$ be the Hauptmodul of  $\Gamma_{\cal N}$? 
If so, all the poles of $Z^{\cal N}$ must be related to  the cusp at $\tau=i\infty$ by the action of $\Gamma_{\cal N}$.  
However, $Z^{\cal N}(\tau)$ diverges at $\tau=0$, 
as can be seen from its $S$ transform $Z_{\cal N}(\tau)$ given in \eqref{ZNx}.  
Hence, for $Z^{\cal N}(\tau)$ to be a Hauptmodul, the full invariance  group must include an element $\gamma \in PSL(2,\mathbb{R})$ that maps $\tau = i \infty$ to $\tau = 0$.  
Such an element $\gamma$ must be of the form 
\ie
\gamma= \left(\begin{array}{cc}0 & -{1\over c} \\c & d\end{array}\right) \in PSL(2,\mathbb{R})\,.
\fe
The condition $Z^{\cal N}(\gamma \cdot \tau ) = Z^{\cal N}(\tau)$ is equivalent to
\ie
Z_{\cal N} (\tau + cd)  = Z^{\cal N} ({\tau\over c^2})\,.
\fe
However, by comparing their explicit expressions \eqref{ZNt} and \eqref{ZNx}, the above equality cannot be true for any $c,d\in \bR$.  
Hence, the defect McKay-Thompson series cannot be the Hauptmodul of  $\Gamma_{\cal N}<PSL(2,\mathbb{R})$.\footnote{For the usual McKay-Thompson series $Z^g$ associated with a group element  $g$ of the Monster group,  $Z^g$ has a pole at $\tau=0$ if and only if it is invariant under the Fricke involution $\tau \to -1/ N\tau$ where $N$ is the order of $g$ \cite{tuite1995}.  
However,  the defect McKay-Thompson series $Z^{\cal N}$ is not invariant under any Fricke involution, and yet it is divergent at $\tau=0$.  }

\subsection{Decomposition into the Baby Monster Representations}

Let us next compute the partition functions of the Monster CFT restricted to different eigenvalues of  $\widehat{\cal N}$.  
From the fusion rule \eqref{fusion}, the zero eigenvalue operators are the $\bZ_{2A}$-odd operators, which are the states in ${\cal H}^-$:
\ie
Z^{\widehat{\cal N}=0}  (\tau )&=Z^{\,\hat\eta=-1}(\tau )= \frac 12 \left[ \, Z(\tau)  -  Z^\eta(\tau) \,\right]  
=      \chi_{1\over16}^{\rm Ising}(\tau )  \chi_{31\over16} ^{\rm Baby}(\tau)\\
&=96256 q + 10698752 q^2 + 431529984 q^3 + 10117578752 q^4+\cdots\,.
\fe
The $\bZ_{2A}$-even states are further split into the  $\widehat{\cal N}=\pm\sqrt{2}$ sectors, which are 
\ie
Z^{ \widehat{\cal N}=+ \sqrt{2} }(\tau )  &= \frac 12\left[\,  Z(\tau) - Z^{\,\hat\eta=-1}(\tau   ) + {1\over \sqrt{2} }Z^{\cal N}(\tau)\,\right]
  =   \chi_0^{\rm Ising}(\tau )  \chi_0 ^{\rm Baby}(\tau)\\
&={1\over q} + 96257 q + 9646892 q^2 + 366941269 q^3 + 8233443176 q^4 +\cdots\,,\\
Z^{  \widehat{\cal N}=- \sqrt{2} }(\tau )  &= \frac 12\left[\,  Z(\tau) - Z^{\,\hat\eta=-1}(\tau   ) - {1\over \sqrt{2} }Z^{\cal N}(\tau)\,\right]
  =   \chi_{1\over2}^{\rm Ising}(\tau )  \chi_{3\over2} ^{\rm Baby}(\tau)   \\
&=4371 q + 1148116 q^2 + 65828717 q^3 + 1894834328 q^4 +\cdots\,.
\fe
For example, at $h=2$, there are 196883 Virasoro primaries transforming in the smallest nontrivial irrep of the Monster group, and a single level-2 descendant of the vacuum, with eigenvalue $+\sqrt{2}$ (because its primary does).  
Out of the 196883 Virasoro primaries, 96256 of them have $\widehat{\cal N}=+\sqrt{2}$, 4371 of them have $\widehat{\cal N}=-\sqrt{2}$, and 96256 of them have $\widehat{\cal N}=0$.  
In particular, the duality defect does \textit{not} commute with the Monster group. 

Interestingly, the first few irreps of   of the Baby Monster group $\mathbb{B}$ are $\bf1$, $\bf 4371$, $\bf 96255$, and the smallest irrep of the double cover $2.\mathbb{B}$ that is not of $\mathbb{B}$ is $\bf 96256$. 
Indeed, the three partition functions $Z^{ \widehat{\cal N}=\pm \sqrt{2},0 } $ can all be written in terms of the Ising and the Baby Monster characters, 
which are known to be decomposable into the $2.\mathbb{B}$   irreps \cite{hoehn2007selbstduale}.  
This generalizes the familiar fact that the coefficients of the $2A$ McKay-Thompson series $Z^\eta$ can be decomposed into the dimensions of the $2.\mathbb{B}$ irreps.

The latter   fact about the $2A$ McKay-Thompson series  $Z^\eta$ can be physically understood as follows.  
Recall that $Z^\eta$ is  the partition function with a $\bZ_{2A}$ twist in the time direction.  
The centralizer of the $\bZ_{2A}$ in the Monster group is $2.\mathbb{B}$, hence the coefficients of $Z^\eta$ can be decomposed into the irreps of $2.\mathbb{B}$ (with signs).  
The same reasoning also applies to the duality defect $\cal N$. 
In the fermionic theory $\cal F$, the chiral fermion parity $\bZ_2^\psi$ commutes with the symmetry group $\mathbb{B}$ of the Baby Monster CFT.  
Since the duality defect $\cal N$ comes from $\bZ_2^\psi$ of $\cal F$ (see Table~\ref{tab:relation}), it commutes with the $2.\mathbb{B}$ symmetry subgroup of the Monster CFT.  
This explains why the coefficients of  $Z^{\widehat{\cal N} =\pm \sqrt{2} ,0}$ can be decomposed into the dimensions of the $2.\mathbb{B}$ irreps.  
 Similarly, the degeneracies in $Z_{\cal N}$ in \eqref{ZNx} can be decomposed into the dimensions of $2.\mathbb{B}$ irreps.

States with $\widehat{\cal N}$ eigenvalues $+\sqrt{2}, \, -\sqrt{2}, \, 0$ are analogous to the identity 1, the energy operator $\varepsilon_{\frac 12, \frac12}$, and the order operator $\sigma_{{1\over 16}  ,{1\over16}}$ in the Ising CFT, respectively. We summarize this analogy in Table~\ref{Tab:Analogy}.\footnote{Note that the 96256 Virasoro primaries at $h=2$ with $\widehat{\cal N}=+\sqrt{2}$ should be decomposed into $\mathbf{1}\oplus\mathbf{96255}$ of $\mathbb{B}$, rather than $\mathbf{96256}$ of $2.\mathbb{B}$, because the Baby Monster module $h=0$ has the $\mathbb{B}$ symmetry and not the $2.\mathbb{B}$.  By contrast, the 96256 Virasoro primaries of the $\widehat{\cal N}=0$ states should be identified with $\mathbf{96256}$ of $2.\mathbb{B}$, which is the symmetry of the Baby Monster module with $h={31\over 16}$.}
\begin{table}[H]
\ie\nonumber
\left.
\renewcommand{\arraystretch}{1.75}%
\begin{array}{|c|c|c|c|}
\hline
& \widehat{\cal N}=+\sqrt{2} & ~~~~\widehat{\cal N}=-\sqrt{2}~~~~ & ~~~~~~~~\widehat{\cal N}=0~~~~~~~~ 
\\\hline
\text{Ising} & 1 & \varepsilon_{\frac 12, \frac12} & \sigma_{{1\over 16}  ,{1\over16}} 
\\\hline
~~~~\text{Monster at $h=2$}~~~~ &~~ T(z), ~\mathbf{1}\oplus\mathbf{96255}~~ & \mathbf{4371} & \mathbf{96256} 
\\
\hline \end{array}
\right.
\fe
\caption{Analogy between the Ising CFT and the Monster CFT, in regards to the duality defect line $\cal N$.}
\label{Tab:Analogy}
\end{table}

\section{General Twisted Partition Functions}
\label{Sec:DefectMcKayThompson}

We can further compute more general torus partition functions twisted by the duality defect $\cal N$ and the $\bZ_{2A}$ line $\eta$.  
We will use several of the $F$-moves of the duality defects and the $\bZ_{2A}$ lines in the Ising category, which we refer the readers to Figure 31 of \cite{Chang:2018iay}.

\begin{figure}[H]
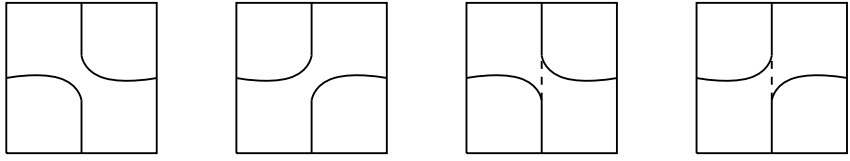

\centering
$\Z{1}{2}{2}{0}{}$ \qquad $\Z{1}{2}{2}{1}{}$
\qquad
$\Z{1}{2}{2}{0}{1}$ \qquad $\Z{1}{2}{2}{1}{1}$
\caption{Four resolutions of $Z_{\cal N}^{\cal N}$.  
The boundary square represents a torus with opposite sides identified. 
The solid line (in the interior of the square) stands for the duality defect $\cal N$, while the dashed line stands for the $\bZ_{2A}$ defect line $\eta$. 
}
\label{fig:ZNNresolution}
\end{figure}

To begin with, consider the partition function  $Z_{\cal N}^{\cal N}$ twisted by $\cal N$ in both the time and the space directions.
As explained in \cite{Chang:2018iay},  partition functions involving an intersection of two topological defect line are generally ambiguous.
To resolve the ambiguity, one needs to resolve the intersection. 
Out of the four possible resolutions shown in Figure \ref{fig:ZNNresolution}, only two of them are linearly independent upon using the $F$-moves of the topological defect lines (see Figure~\ref{fig:Fmoves}).\footnote{Note that there is no ambiguity in $Z_\eta^\eta(\tau) = Z_\eta(\tau\pm1)$ for a non-anomalous $\bZ_2$ symmetry because the $F$-move is trivial.  By contrast, the partition function with $\bZ_2$ twists in both the time and space directions is ambiguous for an anomalous $\bZ_2$ \cite{Lin:2019kpn}. For this reason, we will use the same notation $\hat\eta$ both as a map on $\cal H$ and on the defect Hilbert space ${\cal H}_{\eta}$ twisted by $\bZ_{2A}$.}   
We will choose the two independent ones to be the ones with no intermediate $\eta$ line (the first two configurations in Figure~\ref{fig:ZNNresolution}), and denote the corresponding twisted partition functions by $Z_{\cal N}^{{\cal N}_\pm}$, which can be obtained by performing $T^{-1}$ and $T$  on $Z_{\cal N}$:  
\ie\label{ZNN}
Z_{\cal N}^{{\cal N}_\pm}  (\tau ) &=  Z_{\cal N} (\tau \mp1 )
= e^{ \mp {2\pi i\over 16}}  \,    \chi_{1\over16}^{\rm Ising} \, (\, \chi_0^{\rm Baby} - \chi_{3\over2}^{\rm Baby} ) 
 +  e^{ \pm{ 2\pi i\over 16}} (  \chi_{0}^{\rm Ising} -  \chi_{1\over2}^{\rm Ising} )\,\chi_{31\over16} ^{\rm Baby} \,.
\fe

\begin{figure}[H]
\centering
$Z_{\cal N}^{{\cal N}_+} = \quad \Z{1}{2}{2}{0}{0} \quad = \quad \displaystyle {1\over\sqrt2} \quad
\Z{1}{2}{2}{1}{0}
\quad + \quad {1\over\sqrt2} \quad \Z{1}{2}{2}{1}{1}$
\\
~ \vspace{.25in}
\\
$Z_{\cal N}^{{\cal N}_-} = \quad \Z{1}{2}{2}{1}{0} \quad = \quad \displaystyle {1\over\sqrt2} \quad 
\Z{1}{2}{2}{0}{0}
\quad + {1\over\sqrt2} \quad \Z{1}{2}{2}{0}{1}$
\\
~ \vspace{.25in}
\\
$Z_\eta^{{\cal N}_- } = \quad \Z{1}{2}{1}{1}{} \quad = \quad - \quad \Z{1}{2}{1}{0}{} \quad = - Z_\eta^{{\cal N}_+}$
\\
~ \vspace{.25in}
\\
$Z_{\cal N}^{\eta_- } = \quad \Z{1}{1}{2}{1}{} \quad = \quad - \quad \Z{1}{1}{2}{0}{} \quad = - Z_{\cal N}^{\eta_+}$
\caption{$F$-moves relating the different resolutions for $Z_{\cal N}^{\cal N}$, $Z_{\cal N}^\eta$, and $Z_\eta^{\cal N}$. 
There are two independent resolutions for $Z_{\cal N}^{\cal N}$, and we choose to use the two without an intermediate $\eta$ line. 
 The replacement rule for the other two (the ones with an intermediate $\eta$ line) can be obtained by solving the first two equations.
There is one independent resolution for $Z_{\cal N}^\eta$ and $Z_\eta^{\cal N}$, and we choose to use the ones on the left.}
\label{fig:Fmoves}
\end{figure}

\newpage

Next, we consider the partition function with $\cal N$ extended along the space direction and $\eta$ extended along the time direction.  Again, this twisted partition function has an ambiguity in the intersection between the two lines.  In this case, the $F$-moves of the Ising category imply that they only differ by an overall sign, $Z_{\eta}^{\cal N_+}  = - Z_{\eta}^{\cal N_-}$.  See Figure~\ref{fig:Fmoves}.  
From \eqref{fermionization}, we see that $Z_{\eta}^{\cal N_-}$ comprises exactly of the half-integer power terms of the fermionic partition function $Z_{NS}^\psi$:
\ie\label{ZetaN}
Z_{\eta}^{{\cal N}_-}  (\tau )&  =  \text{Tr}_{ {\cal H}_{\eta}} \left[ \widehat
{\cal N}_-\, q^{h-c_L/24}  \right] = - Z_{\eta}^{{\cal N}_+}  (\tau ) \\
 & =- \sqrt{2}i \times \frac 12 \left[ \, Z_{NS}^\psi(\tau)-Z_{NS}^\psi(\tau+1)\,\right]\\
 & =
-\sqrt{2}i \left(\,
- {1\over \sqrt{q} } +4370 \sqrt{q}+1047488 q^{3/2}+54941824 q^{5/2}+\cdots
\,\right)
\,.
\fe
The overall factor $-\sqrt{2}i$ requires some explanation.  On the ordinary Hilbert space $\cal H$, the duality defect $\cal N$ defines a map $\widehat{\cal N}:{\cal H}\to {\cal H}$ obeying  $\widehat{\cal N}\circ\widehat {\cal N} = I+\hat\eta$.  
Consequently, $\widehat{\cal N}$ has eigenvalues $\pm\sqrt{2}$ and 0.  
However, on the defect Hilbert space ${\cal H}_\eta$ of $\bZ_{2A}$, the fusion is modified in the following sense. 
Let us define the map $\widehat{\cal N}_\pm : {\cal H}_\eta \to {\cal H}_\eta$  on the defect Hilbert space of $\bZ_{2A}$ as in Figure \ref{fig:hats}. 
Following a sequence of $F$-moves as in Figure~\ref{fig:NNdefect}, we find that $\widehat{\cal N}_+ \circ \widehat{\cal N}_+=\widehat{\cal N}_- \circ \widehat{\cal N}_- = -I +\hat\eta$.  
Therefore, the eigenvalues of $\widehat{\cal N}_+$ are $\pm \sqrt{2} i$ and 0.  Between the two choices $\pm \sqrt{2} i$ of the overall factor, we claim that $- \sqrt{2} i$ is the correct choice.  We show this is in the next paragraph.

Applying the modular $S$ transformation on $Z^{ {\cal N}_+}_{\eta} = - Z^{ {\cal N}_-}_{\eta}$ given in \eqref{ZetaN}, we get
\begin{align}\label{ZNeta}
Z_{ {\cal N}}^{\eta_- } (\tau) &=\text{Tr}_{ {\cal H}_{\cal N}} \left[ \hat
{\eta}_-\, q^{h-c_L/24}  \right] = -Z_{ {\cal N}}^{\eta_+ } (\tau)  \notag  \\
&= {1\over \sqrt{2}  } \sqrt{\theta_2(\tau )\over\theta_3(\tau)} Z_{NS}^{\cal F}(\tau)  
- {1\over \sqrt{2}  } \sqrt{\theta_3(\tau )\over\theta_2(\tau)} Z_R^{\cal F}(\tau)  \\
&= {i \over q^{15/16} }  + i q^{{1\over16}}  + 4371 i q^{9/16} - 96256 i q^{15/16} +96257i q^{17/16}  - 96256 i q^{23/16}+\cdots\,. \notag
\end{align}
See Figure~\ref{fig:Fmoves} for the meaning of $Z_{ {\cal N}}^{\eta_-}$ as a resolution of $Z_{ {\cal N}}^\eta$. 
Here $\hat\eta_\pm : {\cal H}_{\cal N} \to {\cal H}_{\cal N}$ are maps  on the defect Hilbert space of $\cal N$ defined as in Figure \ref{fig:hats}.   
Again by the $F$-move, we have $\hat\eta_+ = - \hat\eta_-$.  
It was shown in Section 6.1.1 of \cite{Chang:2018iay} that the $\hat\eta_\pm$ obeys $\hat\eta_+ \circ \hat\eta_+ = \hat\eta_- \circ \hat\eta_- = -1$. 
Furthermore, the $\hat\eta_-$ eigenvalue is determined by the spin $s=h-\bar h$ of the operator, 
\ie
\hat\eta_-  =
\begin{cases}
& +i\,, ~~~  \text{if $s\in {1\over 16}+{\bZ\over2}$}\,,\\
& -i\,, ~~~  \text{if $s\in -{1\over 16}+{\bZ\over2}$}\,.
\end{cases}
\fe
See (6.4) of \cite{Chang:2018iay} with $\epsilon=+1$ for the derivations.  This spin selection rule is indeed consistent with our previously claimed sign choice for $Z_\eta^{{\cal N}_-}$, and would have been inconsistent with the other sign.

\begin{figure}[H]
\centering
\ie\nonumber
& \widehat{\cal N}_+:
\hspace{.5in}
\begin{gathered}
\begin{tikzpicture}[scale = 1]
\Zpre{1}{2}{1}{0}{}
\node at (1,-.5) {$|\varphi\rangle \in {\cal H}_\eta$};
\end{tikzpicture}
\end{gathered}
\hspace{.5in}
\act{1}{2}{1}{0}
\hspace{.5in}
\widehat{\cal N}_-:
\hspace{.5in}
\begin{gathered}
\begin{tikzpicture}[scale = 1]
\Zpre{1}{2}{1}{1}{}
\node at (1,-.5) {$|\varphi\rangle \in {\cal H}_\eta$};
\end{tikzpicture}
\end{gathered}
\hspace{.5in}
\act{1}{2}{1}{1}
\\
& \widehat\eta_+:
\hspace{.5in}
\begin{gathered}
\begin{tikzpicture}[scale = 1]
\Zpre{1}{1}{2}{0}{}
\node at (1,-.5) {$|\varphi\rangle \in {\cal H}_{\cal N}$};
\end{tikzpicture}
\end{gathered}
\hspace{.5in}
\act{1}{1}{2}{0}
\hspace{.5in}
\widehat\eta_-:
\hspace{.5in}
\begin{gathered}
\begin{tikzpicture}[scale = 1]
\Zpre{1}{1}{2}{1}{}
\node at (1,-.5) {$|\varphi\rangle \in {\cal H}_{\cal N}$};
\end{tikzpicture}
\end{gathered}
\hspace{.5in}
\act{1}{1}{2}{1}
\fe
\caption{The duality defect defines two maps $\widehat{N}_\pm :{\cal H}_\eta \rightarrow {\cal H}_\eta$ on the defect Hilbert space of $\eta$ (top two figures).  
Similarly, the $\bZ_{2A}$ defect defines two maps $\hat\eta_\pm :{\cal H}_{\cal N} \rightarrow {\cal H}_{\cal N}$ on the defect Hilbert space of $\cal N$ (bottom two figures).   
For each map, we show the defect line configurations both on the cylinder (left) and on the plane (right), with the two related by the state-operator map. 
  Using the $F$-move in Figure \ref{fig:Fmoves}, we see that $\widehat{\cal N}_+ = -\widehat{\cal N}_-$ and $\hat\eta_+ = -\hat\eta_-$.}\label{fig:hats}
\end{figure}
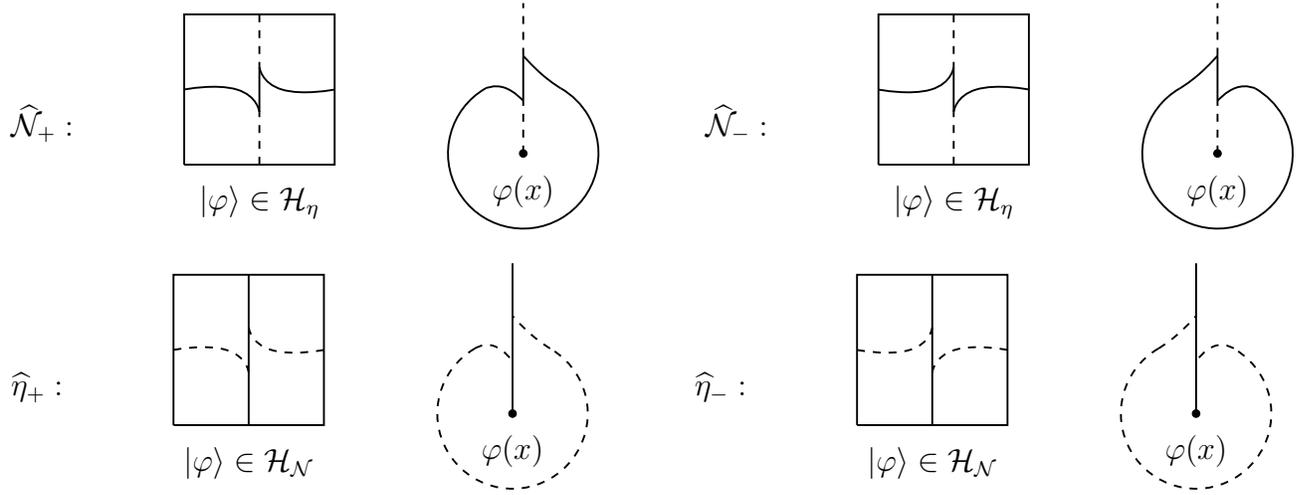

\begin{figure}[H]
\centering
\ie
\nonumber
\begin{gathered}
\begin{tikzpicture}[scale = .67]
\Zthree
	\draw [curve,dashed] plot coordinates {(1.5,0) (1.5,.25) (1,.75) (1,1)};
	\draw [curve,dashed] plot coordinates {(2,1) (2,1.25) (1,1.75) (1,2)};
	\draw [curve,dashed] plot coordinates {(2,2) (2,2.25) (1.5,2.75) (1.5,3)};
\end{tikzpicture}
\end{gathered}
\quad &= \quad - \quad
\begin{gathered}
\begin{tikzpicture}[scale = .67]
\Zthree
	\draw [curve,dashed] plot coordinates {(1.5,0) (1.5,.25) (2,.75) (2,1)};
	\draw [curve,dashed] plot coordinates {(1,1) (1,1.25) (1,1.75) (1,2)};
	\draw [curve,dashed] plot coordinates {(2,2) (2,2.25) (1.5,2.75) (1.5,3)};
\end{tikzpicture}
\end{gathered}
\quad = \quad - \sqrt2   \quad
\begin{gathered}
\begin{tikzpicture}[scale = .67]
\tikzset{line/.style={line width=0.25mm},
curve/.style={line,smooth,tension=1}}
	\draw [line] (0,0) -- (3,0) -- (3,3) -- (0,3) -- (0,0);
	\draw [curve,dashed] plot coordinates {(1.5,0) (1.5,.25) (2,.75) (2,1)};
	\draw [curve,dashed] plot coordinates {(2,2) (2,2.25) (1.5,2.75) (1.5,3)};
	\draw [line] (0,1.5) ++(-90:.5) arc (-90:90:.5);
	\draw [line] (1.5,1.5) ++(90:.5) arc (90:270:.5);
	\draw [line] (1.5,1) -- (3,1);
	\draw [line] (1.5,2) -- (3,2);
\end{tikzpicture}
\end{gathered}
\quad + \quad
\begin{gathered}
\begin{tikzpicture}[scale = .67]
\Zthree
	\draw [curve,dashed] plot coordinates {(1.5,0) (1.5,.25) (2,.75) (2,1)};
	\draw [curve,dashed] plot coordinates {(2,2) (2,2.25) (1.5,2.75) (1.5,3)};
\end{tikzpicture}
\end{gathered}
\\
~ \vspace{.5in}
\\
&= \quad - \quad \Z{1}{}{1}{}{} \quad + \quad \Z{1}{1}{1}{}{}
\fe
\caption{The upper left diagram illustrates the map $\widehat{\cal N}_+\circ \widehat{\cal N}_+$ on the defect Hilbert space ${\cal H}_\eta$ of $\bZ_{2A}$.  By applying a sequence of $F$-moves, we arrive at the second row, which implies that $\widehat{\cal N}_+ \circ \widehat{\cal N}_+ = -I  + \hat\eta$.}
\label{fig:NNdefect}
\end{figure}
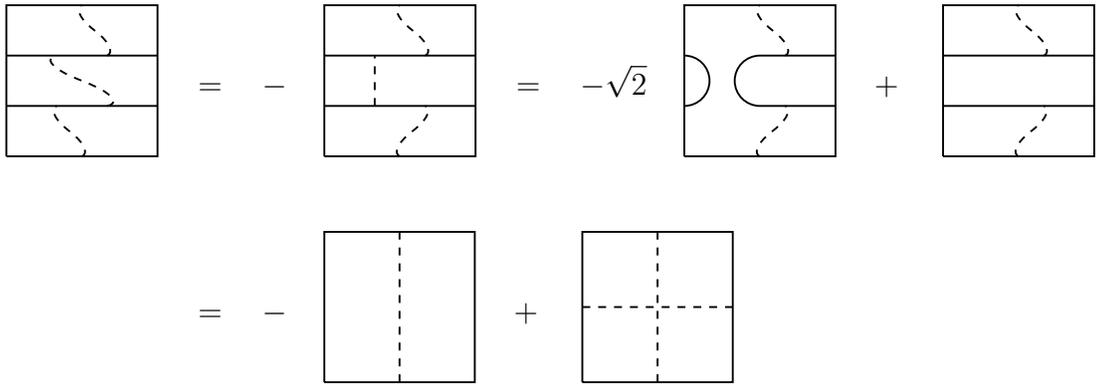

\newpage

To summarize, we have computed all the torus partition functions twisted by the $\bZ_{2A}$ line $\eta$ and the duality defect line $\cal N$:
\ie\label{Monster10}
\Z{.5}{}{}{}{} \qquad &Z = J =  \chi_{0}^{\rm Ising} \chi_{0}^{\rm Baby}  
+   \chi_{1\over2}^{\rm Ising}  \chi_{3\over2}^{\rm Baby}
+ \chi_{1\over16}^{\rm Ising}  \chi_{31\over16}^{\rm Baby} 
\,,\\
\Z{.5}{1}{}{}{} \qquad &Z^\eta = \chi_{0}^{\rm Ising} \chi_{0}^{\rm Baby}  
+   \chi_{1\over2}^{\rm Ising}  \chi_{3\over2}^{\rm Baby}
 - \chi_{1\over16}^{\rm Ising}  \chi_{31\over16}^{\rm Baby} 
\,,~~~\\
\Z{.5}{}{1}{}{} \qquad &Z_\eta  =\chi_{0}^{\rm Ising} \chi_{3\over2}^{\rm Baby}  +\chi_{1\over2}^{\rm Ising} \chi_{0}^{\rm Baby}  
 +\chi_{1\over16}^{\rm Ising}  \chi_{31\over16}^{\rm Baby} 
\,,\\
\Z{.5}{1}{1}{}{} \qquad &Z_\eta^\eta  = - \chi_{0}^{\rm Ising} \chi_{3\over2}^{\rm Baby}   - \chi_{1\over2}^{\rm Ising} \chi_{0}^{\rm Baby}  
 +\chi_{1\over16}^{\rm Ising}  \chi_{31\over16}^{\rm Baby} \,,\\
\Z{.5}{2}{}{}{} \qquad &Z^{\cal N}   
=  \sqrt{2}   \,  \chi_{0}^{\rm Ising} \chi_{0}^{\rm Baby}    - \sqrt{2}   \,  \chi_{1\over2}^{\rm Ising} \chi_{3\over2}^{\rm Baby}  
\,,\\
\Z{.5}{}{2}{}{} \qquad &Z_{\cal N}    =  \chi_{1\over16}^{\rm Ising}  \, (\chi_0^{\rm Baby} + \chi_{3\over2}^{\rm Baby} ) 
+  (  \chi_{0}^{\rm Ising} +  \chi_{1\over2}^{\rm Ising} ) \,\chi_{31\over16} ^{\rm Baby} \,,\\
\begin{gathered} \Z{.5}{2}{2}{0}{} \\ \Z{.5}{2}{2}{1}{} \end{gathered} \qquad &Z_{\cal N}^{{\cal N}_\pm}  
= 
 e^{ \mp {2\pi i\over 16}}  \,    \chi_{1\over16}^{\rm Ising} \, (\, \chi_0^{\rm Baby}- \chi_{3\over2}^{\rm Baby} ) 
 +  e^{ \pm{ 2\pi i\over 16}} (  \chi_{0}^{\rm Ising} -  \chi_{1\over2}^{\rm Ising} )\,\chi_{31\over16} ^{\rm Baby} \,,\\
\Z{.5}{1}{2}{1}{} \qquad &Z_{ {\cal N}}^{\eta_- }  = i \,  \chi_{1\over16}^{\rm Ising} \, (\chi_0^{\rm Baby} + \chi_{3\over2}^{\rm Baby} ) - i \, (  \chi_{0}^{\rm Ising} +  \chi_{1\over2}^{\rm Ising} ) \,\chi_{31\over16} ^{\rm Baby} 
\,,\\
\Z{.5}{2}{1}{1}{} \qquad &Z_{\eta}^{{\cal N}_-} 
=  - \sqrt{2} i  \,  \chi_{0}^{\rm Ising} \chi_{3\over2}^{\rm Baby}  +  \sqrt{2} i  \,  \chi_{1\over2}^{\rm Ising} \chi_{0}^{\rm Baby}
\,.
\fe
Here we write all the partition functions in terms of the Ising and the Baby Monster characters using Table~\ref{BF}.  
Among the 10 partition functions, there is a linear relation: $Z = Z^\eta +Z_\eta +Z^\eta_\eta$, which is the statement that the theory is self-dual under the $\bZ_{2A}$ orbifold.

\newpage

For comparison, we show the corresponding 10 twisted partition functions in the Ising CFT (we drop the superscript ``Ising" below):\footnote{We thank Wenjie Ji and Xiao-Gang Wen for discussions on this point.}
\ie\label{Ising10}
&Z = |\chi_0|^2+|\chi_{1\over2}|^2+|\chi_{1\over16}|^2\,,\\
&Z^\eta = |\chi_0|^2+|\chi_{1\over2}|^2 - |\chi_{1\over16}|^2\,,~~~\\
&Z_\eta =  \chi_0\bar \chi_{1\over2}  +\chi_{1\over2} \bar\chi_0 +|\chi_{1\over16}|^2\,,\\
&Z_\eta^\eta = - \chi_0\bar \chi_{1\over2}  - \chi_{1\over2} \bar\chi_0 +|\chi_{1\over16}|^2\,,\\
&Z^{\cal N} = \sqrt{2}  |\chi_0|^2  -\sqrt{2} |\chi_{1\over2}|^2\,,\\
&Z_{\cal N}   =  \chi_{1\over16} ( \bar\chi_0+\bar \chi_{1\over2})  +    (\chi_0+\chi_{1\over2}) \bar \chi_{1\over16}\,,\\
&Z_{\cal N}^{{\cal N}_\pm}   
= e^{\mp {2\pi i \over 16}} \chi_{1\over16} ( \bar\chi_0 - \bar \chi_{1\over2})  + e^{\pm {2\pi i \over 16}}    (\chi_0- \chi_{1\over2}) \bar \chi_{1\over16} \,,\\
&Z_{ {\cal N}}^{\eta_- } 
= i \,\chi_{1\over16} ( \bar\chi_0+\bar \chi_{1\over2})   - i\,  (\chi_0+\chi_{1\over2}) \bar \chi_{1\over16}\,,\\
&Z_{\eta}^{{\cal N}_-}    
=- \sqrt{2}i \, \chi_0\bar \chi_{1\over2}   +\sqrt{2} i \,\chi_{1\over2} \bar\chi_0 \,,
\fe
where $\bar\chi_{\bar h }   = \chi_{\bar h}(\bar\tau)$.   Note that the Monster twisted partition functions can be obtained from those of the Ising CFT by replacing the right-moving Ising characters $\bar \chi_{0, \, {1\over2}, \, {1\over16} }^{\rm Ising}$ by $\chi^{\rm Baby}_{0, \, {3\over2}, \, {31\over16}}$.  
This reflects the fact that the modular $S$ and $T$ matrices of the Baby Monster modules are the complex conjugates of those  of the Ising CFT (see Appendix~\ref{App:Modular}).

\section{Summary and Outlook}

We end with a summary of our results and several open questions for future exploration.

\subsubsection*{Summary}
\begin{itemize}
\item We showed that the fermionization of the (bosonic) Monster CFT with respect to $\bZ_{2A}$ is the tensor product of a Majorana-Weyl fermion and the (fermionic) Baby Monster CFT.   
\item  Through the above bosonization/fermionization relation, we identified a non-invertible duality defect $\cal N$ in the Monster CFT, which together with the $\bZ_{2A}$ symmetry form an Ising category.  
This Ising category commutes with the $2.\mathbb{B}$ subgroup of the Monster group, and extends the Monster group symmetry to a larger tensor category of topological defect lines.
\item We introduced the defect McKay-Thompson series $Z^{\cal N}$ as the Monster partition function twisted by the duality defect $\cal N$, and showed that its coefficients can be decomposed into the dimensions of the $2.\mathbb{B}$ irreps.  
\item 
The defect McKay-Thompson series $Z^{\cal N}$ is invariant under $\Gamma_0(8)$ up to a sign.  
It is invariant under the congruence subgroup $16D^0$ of $PSL(2,\bZ)$ that is genus zero, index 24, and level 16 .  
We further showed that $Z^{\cal N}$ cannot be the Hauptmodul of its invariance group inside $PSL(2,\mathbb{R})$.  
\end{itemize}

\subsubsection*{Open questions}
\begin{itemize}
\item What is the full set of topological defect lines in the Monster CFT?  
What is the tensor category they form? 
 The fusion subcategory consisting only of the invertible lines, or equivalently, the 't Hooft anomaly of the Monster group was studied in \cite{Johnson-Freyd:2017ble}.
\item Our investigation of $\bZ_{2A}$ has a natural generalization to $\bZ_{pA}$ with $p$ a prime number dividing the order of the Monster group. 
For each such $p$, the Monster CFT is known to be self-dual under the  $\bZ_{pA}$ orbifold \cite{tuite1995,Paquette:2016xoo,Paquette:2017xui,2017arXiv170509022A,2017arXiv170702954C}, which implies that the Monster CFT must have a duality defect belonging to a $\bZ_p$ Tamabara-Yamagami category.\footnote{We thank Theo Johnson-Freyd for pointing this out to us.}   
For each $p$, it would be interesting to determine which one of the two $\bZ_p$ Tamabara-Yamagami categories is realized, perhaps by identifying relations to $\bZ_p$ parafermions and realizing the $\cal N$ line in the parafermion theory.  
\item Conway and Norton \cite{Conway} conjectured that every McKay-Thompson series is the unique Hauptmodul of a genus-zero subgroup of $PSL(2,\mathbb{R})$. 
For example, the $2A$ series is the Hauptmodul of $\Gamma_0(2)+$.  
One could ask whether the defect McKay-Thompson series $Z^{\cal N}$ of Section~\ref{Sec:DefectMcKayThompson} as well as the generalizations to $\bZ_{pA}$ boasts any interesting modular properties.
\item Generally, given a bosonic theory $\cal B$ that is self-duality under a $\bZ_2$ orbifold, {\it i.e.} ${\cal B} = {\cal B}/\bZ_2$, how do we determine the category of the corresponding duality defect $\cal N$?  Which one of the two $\bZ_2$ Tamabara-Yamagami categories ({\it i.e.} the Ising category and the $SU(2)_2$ category) does it realize?  In the Monster CFT, we determined the category to be the Ising category via the free fermion sector in the fermionized theory.  
\item   Can other bilinear relations, such as the ones discussed in \cite{Gaberdiel:2016zke,Hampapura:2016mmz,Harvey:2018rdc,Bae:2018qfh}, be understood in terms of (analogs of) bosonization/fermionization?
\end{itemize}

\section*{Acknowledgements}

We thank Nathan Benjamin, Meng Cheng, Jeffrey Harvey,  Wenjie Ji, Petr Kravchuk, Theo Johnson-Freyd, Natalie Paquette, and Xiao-Gang Wen for discussions. 
SHS would like to thank Kantaro Ohmori and Nathan Seiberg for useful discussions on global symmetries  in bosonic and fermionic QFTs.  
We thank Theo Johnson-Freyd for comments on the draft. 
YL is supported by the Sherman Fairchild Foundation, and by the U.S. Department of Energy, Office of Science, Office of High Energy Physics, under Award Number DE-SC0011632. The work of  SHS is supported by the Simons Foundation/SFARI (651444, NS).

\appendix

\section{Irreducible Representations of the Baby Monster Group}

We denote the Baby Monster group by $\mathbb{B}$, and its double cover by $2.\mathbb{B}$. The latter is a subgroup of the Monster group, while the former is not.  The dimensions of the first few irreps are
\ie
& 1, \quad 4371, \quad 96255, \quad \underline{96256}, \quad 1139374, \quad 9458750, \quad 9550635, \quad \underline{10506240},
\\
& 63532485, \quad 347643114, \quad 356054375, \quad \underline{410132480}, \quad 1407126890, \quad \cdots \,,
\fe
where the underlined ones are irreps of $2.\mathbb{B}$ but not of $\mathbb{B}$, and the rest are irreps of both $2.\mathbb{B}$ and $\mathbb{B}$.

\section{Baby Monster {VOA}}
\label{App:VOA}

The Baby Monster {VOA} with central charge $c = {47\over2}$ was constructed by \cite{hoehn2007selbstduale} using the Monster VOA.  It has three irreducible modules with conformal weights $0$, $1\over2$, and $31\over16$, which we denote by $V^{\rm Baby}_0$, $V^{\rm Baby}_{1\over2}$, and $V^{\rm Baby}_{31\over16}$.  In this appendix we review some key facts about the Baby Monster {VOA} and its irreducible modules.

\subsection{Twisted Modules of the Monster VOA}

The mathematical construction of the Baby Monster {VOA} uses the $\bZ_{2A}$-twisted module of the Monster VOA.  Mathematically, the existence of such a module was part of the Generalized Moonshine Conjecture    \cite{norton1987generalized}.  Physically, this module is the defect Hilbert space of the Monster CFT, where the boundary condition on the space circle is twisted by $\bZ_{2A}$, represented by the insertion of a topological defect line $\eta$ (for $\bZ_{2A}$) {extended} along the time direction.  In both the original or defect Hilbert spaces, the states are be separated into $\bZ_{2A}$-even/odd sectors, and thus we have four sectors ${\cal H}^\pm,  {\cal H}_\eta^\pm$. 
Under OPE, the $\bZ_{2A}$-even operators of ${\cal H}^+$ form a VOA $V_{00}$, and each of the other three sectors forms a $V_{00}$-module, which we denote by $V_{01}$, $V_{10}$, and $V_{11}$, respectively (in the notation of \cite{hoehn2007selbstduale}).  
Due to the self-duality \eqref{selfdual},  we have  $V_{01} \cong V_{10}$ whose primary has conformal weight 2.
Finally, the primary of $V_{11}$ has conformal weight $1\over2$, which is the free Majorana-Weyl fermion.  To make clear the following discussion of the relation to the Baby Monster {VOA}, we denote these modules by $V^{\rm Monster}_0$, $V^{\rm Monster}_2$, and $V^{\rm Monster}_{1\over2}$.

The characters for the various modules defined above have the following closed forms:
\ie
\label{MonsterCharacters}
\chi^{\rm Monster}_0(\tau) &= Z^\eta(\tau) + {1\over2} \left[Z^\eta({\tau\over2}) + Z^\eta({\tau+1\over2})\right] \,,
\\
\chi^{\rm Monster}_2(\tau) &= {1\over2} \left[ Z^\eta({\tau\over2}) + Z^\eta({\tau+1\over2})\right] \,,
\\
\chi^{\rm Monster}_{1\over2}(\tau) &= {1\over2} \left[ Z^\eta({\tau\over2}) - Z^\eta({\tau+1\over2})\right] \,,
\fe
where $Z^\eta$ is the $2A$ McKay-Thompson series given in \eqref{Zetat}. 
The modular invariant torus partition function of the Monster CFT is $J(\tau) = j(\tau) - 744 = \chi^{\rm Monster}_0(\tau) + \chi^{\rm Monster}_2(\tau), $
where $j$ denotes Klein's $j$-invariant.

\subsection{Modules and Characters of the Baby Monster}
\label{App:Characters}

The twisted modules of the Monster VOA have the following decomposition into the modules of the Baby Monster and Ising VOAs \cite{hoehn2007selbstduale}:
\ie
\label{VOADecomposition}
V^{\rm Monster}_0 &= V^{\rm Baby}_0 \otimes V^{\rm Ising}_0 \oplus V^{\rm Baby}_{3\over2} \otimes V^{\rm Ising}_{1\over2} \,,
\\
V^{\rm Monster}_2 &= V^{\rm Baby}_{31\over16} \otimes V^{\rm Ising}_{1\over16} \,,
\\
V^{\rm Monster}_{1\over2} &= V^{\rm Baby}_0 \otimes V^{\rm Ising}_{1\over2} \oplus V^{\rm Baby}_{3\over2} \otimes V^{\rm Ising}_0 \,.
\fe

It is clear from the decomposition \eqref{VOADecomposition} that the Baby Monster characters can be written in terms of the the Monster and Ising characters.  The Ising characters are
\ie
\label{IsingCharacters}
\chi^{\rm Ising}_0(\tau) &= {1\over2} \left[\sqrt{\theta_3(\tau) \over \eta(\tau)} + \sqrt{\theta_4(\tau) \over \eta(\tau)}\right] \,,
\\
\chi^{\rm Ising}_{1\over2}(\tau) &= {1\over2} \left[\sqrt{\theta_3(\tau) \over \eta(\tau)} - \sqrt{\theta_4(\tau) \over \eta(\tau)}\right] \,,
\\
\chi^{\rm Ising}_{1\over16}(\tau) &= \sqrt{\theta_2(\tau) \over 2\eta(\tau)} \,.
\fe
Combining \eqref{VOADecomposition}, \eqref{MonsterCharacters}, and \eqref{IsingCharacters}, we obtain the following expressions:
\ie
\label{BabyCharacters}
\chi^{\rm Baby}_0(\tau) &= {\chi^{\rm Monster}_0(\tau) \chi^{\rm Ising}_0(\tau) - \chi^{\rm Monster}_{1\over2}(\tau) \chi^{\rm Ising}_{1\over2}(\tau) \over [\chi^{\rm Ising}_0(\tau)]^2 - [\chi^{\rm Ising}_{1\over2}(\tau)]^2 } 
\\
&= q^{-{47\over48}} \left( 1+96256 q^2+9646891 q^3+O\left(q^4\right) \right) \,,
\\
\chi^{\rm Baby}_{3\over2}(\tau) &= {\chi^{\rm Monster}_{1\over2}(\tau) \chi^{\rm Ising}_0(\tau) - \chi^{\rm Monster}_0(\tau) \chi^{\rm Ising}_{1\over2}(\tau) \over [\chi^{\rm Ising}_0(\tau)]^2 - [\chi^{\rm Ising}_{1\over2}(\tau)]^2 }
\\
&= q^{{3\over2}-{47\over48}} \left( 4371+1143745 q+64680601 q^2+1829005611
   q^3+O\left(q^4\right) \right) \,,
\\
\chi^{\rm Baby}_{31\over16}(\tau) &= {\chi^{\rm Monster}_2(\tau) \over \chi^{\rm Ising}_{1\over16}(\tau)}
\\
&= q^{{31\over16}-{47\over48}} \left( 96256+10602496 q+420831232 q^2+9685952512 q^3+O\left(q^4\right) \right) \,.
\fe
The $h=0$ and $h={3\over2}$ modules each can be decomposed into  irreps of the Baby Monster group $\mathbb{B}$, while the $h={31\over16}$ module can be decomposed into the projective representations of $\mathbb{B}$ ({\it i.e.} they are irreps of $2.\mathbb{B}$ but not of $\mathbb{B}$) \cite{hoehn2007selbstduale}.

\subsection{Modular Properties}
\label{App:Modular}

The characters for the irreducible modules of the Baby Monster {VOA} form a three-dimensional representation of the modular $SL(2, \bZ)$.  More precisely, if the Ising characters transform under the representation ${\bf R}$ of $SL(2, \bZ)$, then the Baby Monster characters transform in the complex conjugate representation $\bar{\bf R}$.  The $S$ and $T$ matrices for the Ising and the Baby Monster {VOA}s can be written uniformly as
\ie
\label{TS}
T = \begin{pmatrix}
e^{-2\pi i {c\over24}} & 0 & 0
\\
0 & e^{2\pi i (-{c\over24}+{1\over2})} & 0
\\
0 & 0 & e^{2\pi i (-{c\over24}+{c\over8})}
\end{pmatrix}\,,
\quad
S = \begin{pmatrix}
{1\over2} & {1\over2} & {1\over\sqrt2}
\\
{1\over2} & {1\over2} & -{1\over\sqrt2}
\\
{1\over\sqrt2} & -{1\over\sqrt2} & 0
\end{pmatrix}\,,
\fe
with $c = {1\over2}$ for Ising and $c = {47\over2}$ for the Baby Monster. Their representations are complex conjugates because ${47\over2} \equiv -{1\over2} \mod 24$.  The tensor product ${\bf R} \otimes \bar{\bf R}$ contains a singlet that corresponds to the modular invariant partition function $J(\tau) = j(\tau) - 744$ of the Monster CFT.  Indeed, at the module level, the Monster CFT $V^{\rm Monster}_0 + V^{\rm Monster}_2$ is the inner product of the Ising and Baby Monster modules as in \eqref{VOADecomposition}.

\section{``Beauty and the Beast" and the $2B$ Fermionization}
\label{App:BB}

In this paper, we have argued that the $\bZ_{2A}$ fermionization of the Monster CFT is the tensor product of the Baby Monster CFT and the Majorana-Wely fermion.  There is another $\bZ_{2B}$ symmetry of the Monster CFT, whose partition function with a twist in the time direction is the $2B$ McKay-Thompson series
\ie\label{2B}
Z^{2B} (\tau)& = \Tr_{\cal H} \left[  \, \widehat\eta^{2B} \,q^{h-c/24} \right]  ={\eta(\tau )^{24}\over\eta(2\tau)^{24} }   +24 
= \frac 1q +276 q -2048 q^2 + 11202q^3 +{\cal O}(q^4)\,.
\fe
What is the $\bZ_{2B}$ fermionization of the Monster CFT?  

The $\bZ_{2B}$ fermionization turns out to be the  ``Beauty and the Beast" ${\cal N}=1$ SCFT 
constructed in \cite{Dixon:1988qd}.    
In fact, their construction in the context of VOA is exactly the fermionization procedure. 
We will check this at the level of the torus partition function below. 

The modular $S$ transformation of \eqref{2B} gives the partition function of the defect Hilbert space associated with $\bZ_{2B}$:
\ie
Z_{2B}(\tau) &= \Tr_{{\cal H}_{2B}} \left[   \,q^{h-c/24} \right]  
=
2^{12} {\eta(\tau)^{24} \over \eta(\tau/2)^{24}} +24
\notag\\
&= 24 + 4096 q^{1/2} +98304 q +1228800 q^{3/2} +10747904 q^2 +{\cal O}(q^{5/2})\,.
\fe
The partition function with $\bZ_{2B}$ twist both in the time and space directions is $Z_{2B}^{2B} (\tau ) =  Z_{2B} (\tau+1)$. 
The NS and R sector partition functions (with and without the $(-1)^F$ inserted) of the $\bZ_{2B}$ fermionization of the Monster CFT are computed by \eqref{fermionization} to be
\ie
Z_{NS}(\tau )   &= \frac12 \left[ \, 
Z(\tau)  +  Z^{2B}(\tau ) + Z_{2B}(\tau) -  Z_{2B}^{2B}(\tau)
\,\right]\\
&
= {1\over q} + 4096 q^{1/2} + 98580 q + 1228800 q^{3/2} + 10745856 q^2 +\cdots\,,\\
\tilde Z_{NS}(\tau )   &= \frac12 \left[ \, 
Z(\tau)  +  Z^{2B}(\tau ) - Z_{2B}(\tau) +  Z_{2B}^{2B}(\tau)
\,\right]\\
&
= {1\over q} - 4096 q^{1/2} + 98580 q - 1228800 q^{3/2} + 10745856 q^2 +\cdots\,,
\\
Z_R(\tau )&  =\frac 12\left[\,
Z(\tau)  -  Z^{2B}(\tau ) + Z_{2B}(\tau) +Z_{2B}^{2B}(\tau)
\,\right]\\
&
=24 + 196608 q + 21495808 q^2 + 864288768 q^3 + 20245905408 q^4+\cdots\,,\\
\tilde Z_R(\tau )&  =\frac 12\left[\,
Z(\tau)  -  Z^{2B}(\tau ) - Z_{2B}(\tau) - Z_{2B}^{2B}(\tau)
\,\right]\\
&
= -24\,.
\fe
These are indeed the spectra of operators in  \cite{Dixon:1988qd}. 
Note that $\tilde Z_R = \Tr_R [(-1)^F q^{h-c/24} ]$ is the Witten index of the ${\cal N}=1$ SCFT.  

Finally, we note that the $\bZ_{2B}$ orbifold of the Monster CFT gives back the Leech lattice CFT:
\ie
Z_{\rm Leech} (\tau )& =   \frac 12\left[\,
Z(\tau)+ Z^{2B}(\tau ) +Z_{2B}(\tau) +Z_{2B}^{2B}(\tau)
\,\right]  \\
&={1\over q} + 24 + 196884 q + 21493760 q^2 + 864299970 q^3  +\cdots\,.
\fe

\section{The Invariance Group of $Z^{\cal N}$ in $\Gamma_0(8)$}
\label{App:16D} 

In the beginning of Section~\ref{Sec:Mod}, we established that $Z^{\cal N}$ transforms under $\Gamma_0(8)$ only by $\pm 1$.
As discussed there, the invariance group $\widetilde \Gamma_{\cal N}$ of $Z^{\cal N}$ in $\Gamma_0(8)$ is the set of words composed of $T, G_1, G_2$ in \eqref{TGG} with an even number of $G_1$.\footnote{Note that the symbol $\Gamma_{\cal N}$ in Section~\ref{Sec:Mod} refers to the invariance group in $PSL(2, \bR)$.  Of course, $\widetilde \Gamma_{\cal N} \le \Gamma_{\cal N}$.
}  
From this definition, it is clear that $\widetilde \Gamma_{\cal N}$ is an index-2 normal subgroup inside $\Gamma_0(8)$.

Now, there is an index-4 normal subgroup 
\ie
\Gamma^{0,0} \equiv \left\{ \left(\begin{array}{cc}a& b \\ c& d\end{array}\right) \in PSL(2,\mathbb{Z}) ~\Big|~ 
a=d=\pm1 ~\text{mod}~8\,,~~c=0~\text{mod}~16
\right\} \,
\fe
inside $\Gamma_0(8)$.\footnote{$\Gamma^{0,0} = \Gamma_0(16) \cap \Gamma_1(8)$, which is called 16$H^0$ in \cite{pauli}.  
}
The four cosets of $\Gamma^{0,0}$ in $\Gamma_0(8)$ are $\Gamma^{0,0}$ and
\ie
& \Gamma^{1,0} \equiv & \left\{ \left(\begin{array}{cc}a& b \\ c& d\end{array}\right) \in PSL(2,\mathbb{Z}) ~\Big|~ 
a=d=\pm3 ~\text{mod}~8\,,~~c=8~\text{mod}~16
\right\} \, ,
\\
& \Gamma^{0,1} \equiv & \left\{ \left(\begin{array}{cc}a& b \\ c& d\end{array}\right) \in PSL(2,\mathbb{Z}) ~\Big|~ 
a=d=\pm3 ~\text{mod}~8\,,~~c=0~\text{mod}~16
\right\} \, ,
\\
& \Gamma^{1,1} \equiv & \left\{ \left(\begin{array}{cc}a& b \\ c& d\end{array}\right) \in PSL(2,\mathbb{Z}) ~\Big|~ 
a=d=\pm1 ~\text{mod}~8\,,~~c=8~\text{mod}~16
\right\} \, ,
\fe
with the quotient group being the Klein four group $\bZ_2 \times \bZ_2$,
\ie
\Gamma^{a,b} \Gamma^{c,d} = \Gamma^{(a+c),(b+d)} \, ,
\fe
where the superscripts are defined modulo 2.

Since $T \in \Gamma^{0,0}$, $G_1 \in \Gamma^{0,1}$, $G_2 \in \Gamma^{1,0}$, it is clear that $\widetilde\Gamma_{\cal N} \le \Gamma^{0,0} \cup \Gamma^{1,0}$.  
But since $\widetilde\Gamma_{\cal N}$ is an index-2 normal subgroup inside $\Gamma_0(8)$, it must be that $\widetilde\Gamma_{\cal N}= \Gamma^{0,0} \cup \Gamma^{1,0}$.  Hence we have shown that $\widetilde\Gamma_{\cal N}$ indeed has the description \eqref{16D0}.

%

\bibliographystyle{JHEP}
\bibliography{KWMonster}

\end{document}